\documentstyle[12pt,bezier]{article}

\newcommand{\densityw}{20}

\newcommand{\be}{\begin{equation}}
\newcommand{\ee}{\end{equation}}
\newcommand{\bea}{\begin{eqnarray}}
\newcommand{\eea}{\end{eqnarray}}
\newcommand{\ep}{\varepsilon}

\newcommand{\Li}[2]{{\mbox{Li}}_{#1}\left(#2\right)}

\begin{document}
\begin{titlepage}
\title{
\begin{flushright} {\large INLO-PUB--5/93} \end{flushright}
 \begin{picture}(150,30) \end{picture}\\
    Large momentum expansion\\[1mm]
    of two-loop self-energy diagrams \\[1mm]
    with arbitrary masses\\[3mm]}
\vspace{10 mm}
\author{ A. I. Davydychev\thanks{ Permanent address: Nuclear Physics
                                 Institute,
                                Moscow State Univ., 119899, Moscow, Russia.
                                E-mail addresses: davyd@compnet.msu.su
                                and smirnov@compnet.msu.su}
                      $\; ^{(a)} \;$ ,  \
         V. A. Smirnov$^* \; ^{(b)} \;$  \
and $\;$ J. B. Tausk\thanks{Research supported by the Stichting FOM.
                           E-mail address: tausk@rulgm0.leidenuniv.nl}
                      $\; ^{(a)} \;$ \\[1cm]
${}^{(a)} \hspace{3mm} $ Instituut--Lorentz,
\ \ \ University \ of \ Leiden,\hspace{5mm} \\
P.O.B. 9506, 2300 RA Leiden, The Netherlands \\[0.3cm]
${}^{(b)}  \hspace{3mm}$
Institut f\"ur Theoretische Physik, Universit\"at G\"ottingen \\
Bunsenstrasse 9, D-3400 G\"ottingen, Germany  \\[1cm]
}

\vskip 15mm

\date{February 1993}
\maketitle

\vspace{9 mm}

\begin{abstract}
\normalsize
{For two-loop two-point diagrams with arbitrary masses, an algorithm to
derive the asymptotic expansion at large external momentum squared is
constructed. By using a general theorem on asymptotic expansions of
Feynman diagrams, the coefficients of the expansion are calculated
analytically.  For some two-loop diagrams occurring in the Standard
Model, comparison with results of numerical integration shows that
our expansion works well in the region above the highest physical
threshold.}
\end{abstract}
\end{titlepage}

\setcounter{page}{2}
\section{Introduction}

The high precision that has been achieved in high energy $e^{+} e^{-}$
experiments, especially at LEP, makes them sensitive to radiative
corrections in the Standard Model at the one and two loop level. However,
some complicated theoretical problems are connected with the
evaluation of two-loop corrections involving massive internal particles
(heavy quarks, $W$ and $Z$ bosons, Higgs particles) whose masses cannot be
neglected in the region of interest.
In particular, much attention has been paid to the evaluation
of massive two-loop self-energy diagrams. In some well known
special cases, such as the QED correction to the photon self-energy,
they can be evaluated exactly \cite{KS,Bro}, and the result can be expressed
in terms
of trilogarithms. The problem becomes essentially more difficult when
all the internal particles of the diagram are massive. Exact results
for the corresponding Feynman integrals are not known, and we need to
look for other approaches to calculate the contributions of such
diagrams.

One of the possible ways is by numerical integration. For this purpose,
it is convenient to use a two-fold integral representation \cite{Kre} obtained
for the so-called ``master'' two-loop two-point integral.
Another way is by constructing asymptotic expansions of such diagrams
in different regions, the coefficients of the expansions being calculated
analytically. For example, in ref.~\cite{DT} we constructed an expansion
for the general case of the two-loop self-energy diagram that works when
the external momentum squared ($k^2$) is below the first physical threshold.
We showed that, unless we are very close to the threshold, only
a few terms of the expansion are needed to obtain highly accurate results.

In the present paper we shall consider another situation, when $k^2$ is
larger than the highest threshold of the diagram. This case is more
complicated, because the corresponding expansion is not a usual Taylor
expansion, but also contains logarithms and squared logarithms of
$(-k^2)$ (in four dimensions) yielding an imaginary part when the
momentum is time-like.  The procedure of calculating this diagram in
the form of a series in inverse powers of the external momentum (plus
logarithms) resembles the standard procedure of analytic continuation
of the hypergeometric function (some explicit examples of such a
procedure connected with Feynman integrals were presented e.g. in
ref.~\cite{BD}). To obtain this expansion, we shall apply a general
mathematical theorem on asymptotic expansions of Feynman integrals in
the limit of large momenta.  This theorem holds at least in the case
when the external momenta are not restricted to a mass shell. However,
in our case of a two-point massive diagram, it is valid for any values
of the external momentum.  All the coefficients of the expansion have a
natural interpretation in terms of Feynman integrals and are
analytically calculable in the case we consider.  Expansions of this
kind were presented in refs.~\cite{PT,C,G,MPI} and rigorously proved in
ref.~\cite{Smi} (see also ref.~\cite{Smi-book} for a review).

The remainder of this paper is organized as follows. In section 2 we explain
how the general theorem can be applied to construct the large momentum
expansion
for the general mass case of two-loop self-energies. We present an algorithm
by which any term of this expansion can be calculated analytically. In section
3
we discuss the analytical results obtained for the coefficients. Section 4
is devoted to a comparison of the asymptotic expansion with the results of
a numerical integration using the method of ref.~\cite{Kre}. Section 5
contains our conclusions. Finally, a number
of formulae we need for massless two-point integrals and for massive
vacuum integrals are collected in appendices A and B, respectively.

\section{Constructing the asymptotic expansion}

We shall consider the ``master'' two-loop self-energy diagram
presented in Fig.1a.
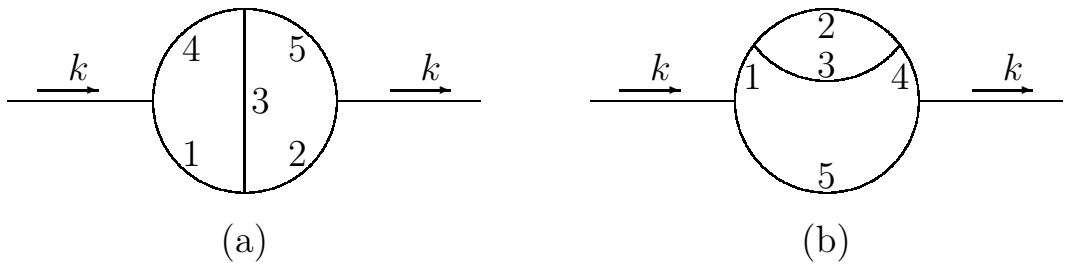
\begin{figure}[b]
\setlength{\unitlength}{0.2mm}
\raisebox{-61.2\unitlength}{
\put(20.0,0.0){
\begin{picture}(314.3,190)( 0.0,-120)
\put( 0.0, 0.0){\line(1,0){96.0}}
\bezier{\densityw}(96.0, 0.0)(96.0,-4.0)(96.5,-8.0)
\bezier{\densityw}(96.5,-8.0)(97.0,-12.0)(98.1,-15.8)
\bezier{\densityw}(98.1,-15.8)(99.1,-19.7)(100.7,-23.4)
\bezier{\densityw}(100.7,-23.4)(102.2,-27.1)(104.2,-30.6)
\bezier{\densityw}(104.2,-30.6)(106.2,-34.1)(108.6,-37.2)
\bezier{\densityw}(108.6,-37.2)(111.1,-40.4)(113.9,-43.2)
\bezier{\densityw}(113.9,-43.2)(116.7,-46.1)(119.9,-48.5)
\bezier{\densityw}(119.9,-48.5)(123.1,-51.0)(126.6,-53.0)
\bezier{\densityw}(126.6,-53.0)(130.1,-55.0)(133.8,-56.5)
\bezier{\densityw}(133.8,-56.5)(137.5,-58.0)(141.3,-59.1)
\bezier{\densityw}(141.3,-59.1)(145.2,-60.1)(149.2,-60.6)
\bezier{\densityw}(149.2,-60.6)(153.2,-61.2)(157.2,-61.2)
\bezier{\densityw}(157.2,-61.2)(161.2,-61.2)(165.1,-60.6)
\bezier{\densityw}(165.1,-60.6)(169.1,-60.1)(173.0,-59.1)
\bezier{\densityw}(173.0,-59.1)(176.9,-58.0)(180.6,-56.5)
\bezier{\densityw}(180.6,-56.5)(184.3,-55.0)(187.7,-53.0)
\bezier{\densityw}(187.7,-53.0)(191.2,-51.0)(194.4,-48.5)
\bezier{\densityw}(194.4,-48.5)(197.6,-46.1)(200.4,-43.2)
\bezier{\densityw}(200.4,-43.2)(203.2,-40.4)(205.7,-37.2)
\bezier{\densityw}(205.7,-37.2)(208.1,-34.1)(210.1,-30.6)
\bezier{\densityw}(210.1,-30.6)(212.1,-27.1)(213.7,-23.4)
\bezier{\densityw}(213.7,-23.4)(215.2,-19.7)(216.2,-15.8)
\bezier{\densityw}(216.2,-15.8)(217.3,-12.0)(217.8,-8.0)
\bezier{\densityw}(217.8,-8.0)(218.3,-4.0)(218.3, 0.0)
\bezier{\densityw}(218.3, 0.0)(218.3, 4.0)(217.8, 8.0)
\bezier{\densityw}(217.8, 8.0)(217.3,12.0)(216.2,15.8)
\bezier{\densityw}(216.2,15.8)(215.2,19.7)(213.7,23.4)
\bezier{\densityw}(213.7,23.4)(212.1,27.1)(210.1,30.6)
\bezier{\densityw}(210.1,30.6)(208.1,34.1)(205.7,37.2)
\bezier{\densityw}(205.7,37.2)(203.2,40.4)(200.4,43.2)
\bezier{\densityw}(200.4,43.2)(197.6,46.1)(194.4,48.5)
\bezier{\densityw}(194.4,48.5)(191.2,51.0)(187.7,53.0)
\bezier{\densityw}(187.7,53.0)(184.3,55.0)(180.6,56.5)
\bezier{\densityw}(180.6,56.5)(176.9,58.0)(173.0,59.1)
\bezier{\densityw}(173.0,59.1)(169.1,60.1)(165.1,60.6)
\bezier{\densityw}(165.1,60.6)(161.2,61.2)(157.2,61.2)
\put(218.3, 0.0){\line(1,0){96.0}}
\bezier{\densityw}(157.2,61.2)(153.2,61.2)(149.2,60.6)
\bezier{\densityw}(149.2,60.6)(145.2,60.1)(141.3,59.1)
\bezier{\densityw}(141.3,59.1)(137.5,58.0)(133.8,56.5)
\bezier{\densityw}(133.8,56.5)(130.1,55.0)(126.6,53.0)
\bezier{\densityw}(126.6,53.0)(123.1,51.0)(119.9,48.5)
\bezier{\densityw}(119.9,48.5)(116.7,46.1)(113.9,43.2)
\bezier{\densityw}(113.9,43.2)(111.1,40.4)(108.6,37.2)
\bezier{\densityw}(108.6,37.2)(106.2,34.1)(104.2,30.6)
\bezier{\densityw}(104.2,30.6)(102.2,27.1)(100.7,23.4)
\bezier{\densityw}(100.7,23.4)(99.1,19.7)(98.1,15.8)
\bezier{\densityw}(98.1,15.8)(97.0,12.0)(96.5, 8.0)
\bezier{\densityw}(96.5, 8.0)(96.0, 4.0)(96.0, 0.0)
\put(157.2,61.2){\line(0,-1){122.3}}
\put (   20,  7){\vector(1, 0){40}}
\put (254.3,  7){\vector(1, 0){40}}
\put (   40, 14){\makebox(0,0)[bl]{\large $k$}}
\put (274.3, 14){\makebox(0,0)[bl]{\large $k$}}
\put (  122, 35){\makebox(0,0)[c]{\large $4$}}
\put (  122,-35){\makebox(0,0)[c]{\large $1$}}
\put (192.5, 35){\makebox(0,0)[c]{\large $5$}}
\put (192.5,-35){\makebox(0,0)[c]{\large $2$}}
\put (  168,  0){\makebox(0,0)[c]{\large $3$}}
\put (157,-94){\makebox(0,0)[c]{\large ${\mbox{(a)}}$}}
\end{picture}
}}
\raisebox{-61.2\unitlength}{
\put(400.0,0.0){
\begin{picture}(314.3,190)( 0.0,-120)
\put( 0.0, 0.0){\line(1,0){96.0}}
\bezier{\densityw}(96.0, 0.0)(96.0,-4.0)(96.5,-8.0)
\bezier{\densityw}(96.5,-8.0)(97.0,-12.0)(98.1,-15.8)
\bezier{\densityw}(98.1,-15.8)(99.1,-19.7)(100.7,-23.4)
\bezier{\densityw}(100.7,-23.4)(102.2,-27.1)(104.2,-30.6)
\bezier{\densityw}(104.2,-30.6)(106.2,-34.1)(108.6,-37.2)
\bezier{\densityw}(108.6,-37.2)(111.1,-40.4)(113.9,-43.2)
\bezier{\densityw}(113.9,-43.2)(116.7,-46.1)(119.9,-48.5)
\bezier{\densityw}(119.9,-48.5)(123.1,-51.0)(126.6,-53.0)
\bezier{\densityw}(126.6,-53.0)(130.1,-55.0)(133.8,-56.5)
\bezier{\densityw}(133.8,-56.5)(137.5,-58.0)(141.3,-59.1)
\bezier{\densityw}(141.3,-59.1)(145.2,-60.1)(149.2,-60.6)
\bezier{\densityw}(149.2,-60.6)(153.2,-61.2)(157.2,-61.2)
\bezier{\densityw}(157.2,-61.2)(161.2,-61.2)(165.1,-60.6)
\bezier{\densityw}(165.1,-60.6)(169.1,-60.1)(173.0,-59.1)
\bezier{\densityw}(173.0,-59.1)(176.9,-58.0)(180.6,-56.5)
\bezier{\densityw}(180.6,-56.5)(184.3,-55.0)(187.7,-53.0)
\bezier{\densityw}(187.7,-53.0)(191.2,-51.0)(194.4,-48.5)
\bezier{\densityw}(194.4,-48.5)(197.6,-46.1)(200.4,-43.2)
\bezier{\densityw}(200.4,-43.2)(203.2,-40.4)(205.7,-37.2)
\bezier{\densityw}(205.7,-37.2)(208.1,-34.1)(210.1,-30.6)
\bezier{\densityw}(210.1,-30.6)(212.1,-27.1)(213.7,-23.4)
\bezier{\densityw}(213.7,-23.4)(215.2,-19.7)(216.2,-15.8)
\bezier{\densityw}(216.2,-15.8)(217.3,-12.0)(217.8,-8.0)
\bezier{\densityw}(217.8,-8.0)(218.3,-4.0)(218.3, 0.0)
\put(218.3, 0.0){\line(1,0){96.0}}
\bezier{\densityw}(96.0, 0.0)(96.0, 4.0)(96.5, 8.0)
\bezier{\densityw}(96.5, 8.0)(97.0,12.0)(98.1,15.8)
\bezier{\densityw}(98.1,15.8)(99.1,19.7)(100.7,23.4)
\bezier{\densityw}(100.7,23.4)(102.2,27.1)(104.2,30.6)
\bezier{\densityw}(104.2,30.6)(106.2,34.1)(108.6,37.2)
\bezier{\densityw}(218.3, 0.0)(218.3, 4.0)(217.8, 8.0)
\bezier{\densityw}(217.8, 8.0)(217.3,12.0)(216.2,15.8)
\bezier{\densityw}(216.2,15.8)(215.2,19.7)(213.7,23.4)
\bezier{\densityw}(213.7,23.4)(212.1,27.1)(210.1,30.6)
\bezier{\densityw}(210.1,30.6)(208.1,34.1)(205.7,37.2)
\bezier{\densityw}(108.6,37.2)(110.9,34.3)(113.5,31.6)
\bezier{\densityw}(113.5,31.6)(116.2,28.9)(119.1,26.6)
\bezier{\densityw}(119.1,26.6)(122.0,24.3)(125.2,22.3)
\bezier{\densityw}(125.2,22.3)(128.4,20.4)(131.8,18.8)
\bezier{\densityw}(131.8,18.8)(135.2,17.3)(138.8,16.1)
\bezier{\densityw}(138.8,16.1)(142.3,15.0)(146.0,14.3)
\bezier{\densityw}(146.0,14.3)(149.7,13.6)(153.4,13.4)
\bezier{\densityw}(153.4,13.4)(157.2,13.2)(160.9,13.4)
\bezier{\densityw}(160.9,13.4)(164.6,13.6)(168.3,14.3)
\bezier{\densityw}(168.3,14.3)(172.0,15.0)(175.6,16.1)
\bezier{\densityw}(175.6,16.1)(179.1,17.3)(182.5,18.8)
\bezier{\densityw}(182.5,18.8)(185.9,20.4)(189.1,22.3)
\bezier{\densityw}(189.1,22.3)(192.3,24.3)(195.2,26.6)
\bezier{\densityw}(195.2,26.6)(198.2,28.9)(200.8,31.6)
\bezier{\densityw}(200.8,31.6)(203.4,34.3)(205.7,37.2)
\bezier{\densityw}(108.6,37.2)(110.9,40.2)(113.5,42.9)
\bezier{\densityw}(113.5,42.9)(116.2,45.5)(119.1,47.9)
\bezier{\densityw}(119.1,47.9)(122.0,50.2)(125.2,52.1)
\bezier{\densityw}(125.2,52.1)(128.4,54.1)(131.8,55.7)
\bezier{\densityw}(131.8,55.7)(135.2,57.2)(138.8,58.3)
\bezier{\densityw}(138.8,58.3)(142.3,59.5)(146.0,60.1)
\bezier{\densityw}(146.0,60.1)(149.7,60.8)(153.4,61.0)
\bezier{\densityw}(153.4,61.0)(157.2,61.3)(160.9,61.0)
\bezier{\densityw}(160.9,61.0)(164.6,60.8)(168.3,60.1)
\bezier{\densityw}(168.3,60.1)(172.0,59.5)(175.6,58.3)
\bezier{\densityw}(175.6,58.3)(179.1,57.2)(182.5,55.7)
\bezier{\densityw}(182.5,55.7)(185.9,54.1)(189.1,52.1)
\bezier{\densityw}(189.1,52.1)(192.3,50.2)(195.2,47.9)
\bezier{\densityw}(195.2,47.9)(198.2,45.5)(200.8,42.9)
\bezier{\densityw}(200.8,42.9)(203.4,40.2)(205.7,37.2)
\put (   20,  7){\vector(1, 0){40}}
\put (254.3,  7){\vector(1, 0){40}}
\put (   40, 14){\makebox(0,0)[bl]{\large $k$}}
\put (274.3, 14){\makebox(0,0)[bl]{\large $k$}}
\put (  108, 16){\makebox(0,0)[c]{\large $1$}}
\put (157.2, 50){\makebox(0,0)[c]{\large $2$}}
\put (157.2, 24){\makebox(0,0)[c]{\large $3$}}
\put (206.3, 16){\makebox(0,0)[c]{\large $4$}}
\put (157.2,-50){\makebox(0,0)[c]{\large $5$}}
\put (157,-94){\makebox(0,0)[c]{\large ${\mbox{(b)}}$}}
\end{picture}
}}
\caption{Two-loop self-energy diagrams}
\end{figure}
The corresponding Feynman integral can be
defined as
\be
\label{defJ}
 J(\{\nu_i\} ; \{m_i\} ; k) = \int \int
\frac{\mbox{d}^n p \; \mbox{d}^n q}
     {D_1^{\nu_1} D_2^{\nu_2} D_3^{\nu_3} D_4^{\nu_4} D_5^{\nu_5} } ,
\ee
where $D_i = (p_i^2-m_i^2+i0)$ are massive denominators (for brevity, we shall
omit the ``causal'' $i0$'s below), $\nu_i$ are
the powers of these denominators, $n=4-2\ep$ is the space-time
dimension (in the framework of dimensional regularization \cite{tHV-72,BolGi}),
and the momenta of the lines $p_i$ are constructed from the external momentum
$k$ and the loop integration momenta $p$ and $q$ (with due account of the
momentum conservation in all vertices). Two-loop diagrams with three or
four internal lines correspond to the cases when some of the $\nu$'s in
eq.~(\ref{defJ}) are equal to zero. Moreover, in the case of integer
$\nu$'s the two-loop diagram with another topology (see Fig.1b)
can also be reduced to integrals (\ref{defJ}), if we use the
decomposition of first and fourth denominators (see e.g. in \cite{DT}).
So, if we deal with
integer powers of denominators, it is in general sufficient to consider
the integrals (\ref{defJ}) with different powers $\nu_i$.

Now let us introduce the result that the asymptotic expansion theorem
gives for our case (the general case of this theorem can be found
in ref.~\cite{Smi-book}):
\be
\label{theorem}
J_{\Gamma} \begin{array}{c} \frac{}{}  \\
                    {\mbox{\Huge$\sim$}} \\ {}_{k^2 \to \infty}
            \end{array}
\sum_{\gamma} J_{\Gamma/\gamma} \; \circ  \;
{\cal{T}}_{\{m_i\}; \{q_i\}} J_{\gamma} ,
\ee
where $\Gamma$ is the main graph (see Fig.1a), $\gamma$ are subgraphs involved
in the asymptotic expansion (see below), $\Gamma/\gamma$ is the reduced graph
obtained from $\Gamma$ by shrinking the subgraph $\gamma$ to a single point,
$J_{\gamma}$ denotes
the dimensionally-regularized Feynman integral corresponding to a graph
$\gamma$ (for example, $J_{\Gamma}$ is given by (\ref{defJ})),
${\cal{T}}_{\{m_i\}; \{q_i\}}$ is the operator of Taylor expansion of the
integrand in masses and ``small'' momenta $q_i$ (that are ``external''
for the given subgraph $\gamma$, but do not contain the ``large'' external
momentum $k$), and the symbol ``$\circ$'' means that the resulting
polynomial in these momenta should be inserted into the numerator of
the integrand of $J_{\Gamma/\gamma}$. It is implied that the
operator $\cal{T}$ acts on the integrands before the loop integrations
are performed.

In our case, the sum (\ref{theorem}) goes over all
subgraphs $\gamma$ that become one-particle irreducible when we
connect the two vertices with external momentum $k$ by a line. In other
words, we should consider all possibilities to ``distribute'' the
external momentum $k$ among $p_i$, and for each arrangement the subgraph
$\gamma$
will coincide with the subset of lines involving $k$, while lines
without $k$ should be removed. For our graph $\Gamma$ (see Fig.1a),
all possible subgraphs $\gamma$ (there are five different types of
them) are presented in Fig.2 (dotted lines correspond to the lines
that do not belong to $\gamma$).
\newcommand{\suba}{
\setlength{\unitlength}{0.253mm}
\raisebox{-15.5\unitlength}{
\begin{picture}(78.9,30.9)( 0.0,-15.5)
\put( 0.0, 0.0){\line(1,0){24.0}}
\bezier{\densityw}(24.0, 0.0)(24.0,-4.1)(26.1,-7.7)
\bezier{\densityw}(26.1,-7.7)(28.1,-11.3)(31.7,-13.4)
\bezier{\densityw}(31.7,-13.4)(35.3,-15.5)(39.5,-15.5)
\bezier{\densityw}(39.5,-15.5)(43.6,-15.5)(47.2,-13.4)
\bezier{\densityw}(47.2,-13.4)(50.8,-11.3)(52.8,-7.7)
\bezier{\densityw}(52.8,-7.7)(54.9,-4.1)(54.9, 0.0)
\bezier{\densityw}(54.9, 0.0)(54.9, 4.1)(52.8, 7.7)
\bezier{\densityw}(52.8, 7.7)(50.8,11.3)(47.2,13.4)
\bezier{\densityw}(47.2,13.4)(43.6,15.5)(39.5,15.5)
\put(54.9, 0.0){\line(1,0){24.0}}
\bezier{\densityw}(39.5,15.5)(35.3,15.5)(31.7,13.4)
\bezier{\densityw}(31.7,13.4)(28.1,11.3)(26.1, 7.7)
\bezier{\densityw}(26.1, 7.7)(24.0, 4.1)(24.0, 0.0)
\put(39.5,15.5){\line(0,-1){30.9}}
\end{picture}
} }
\newcommand{\subb}{
\setlength{\unitlength}{0.069mm}
\raisebox{-56.1\unitlength}{
\begin{picture}(288.2,112.1)( 0.0,-56.1)
\put( 0.0, 0.0){\line(1,0){88.0}}
\bezier{\densityw}(88.0, 0.0)(88.0,-0.4)(88.0,-0.9)
\bezier{\densityw}(88.7,-8.8)(88.8,-9.2)(88.8,-9.6)
\bezier{\densityw}(90.7,-17.3)(90.9,-17.7)(91.0,-18.2)
\bezier{\densityw}(94.1,-25.5)(94.3,-25.9)(94.5,-26.2)
\bezier{\densityw}(98.7,-33.0)(99.0,-33.3)(99.2,-33.7)
\bezier{\densityw}(104.4,-39.7)(104.7,-40.0)(105.1,-40.3)
\bezier{\densityw}(111.1,-45.4)(111.5,-45.6)(111.8,-45.9)
\bezier{\densityw}(118.6,-50.0)(119.0,-50.2)(119.4,-50.4)
\bezier{\densityw}(126.8,-53.3)(127.2,-53.5)(127.6,-53.6)
\bezier{\densityw}(135.3,-55.4)(135.7,-55.5)(136.2,-55.5)
\bezier{\densityw}(144.1,-56.1)(148.1,-56.1)(152.1,-55.5)
\bezier{\densityw}(152.1,-55.5)(156.0,-54.9)(159.9,-53.8)
\bezier{\densityw}(159.9,-53.8)(163.7,-52.7)(167.4,-51.0)
\bezier{\densityw}(167.4,-51.0)(171.0,-49.3)(174.4,-47.2)
\bezier{\densityw}(174.4,-47.2)(177.8,-45.0)(180.8,-42.4)
\bezier{\densityw}(180.8,-42.4)(183.8,-39.8)(186.5,-36.7)
\bezier{\densityw}(186.5,-36.7)(189.1,-33.7)(191.2,-30.3)
\bezier{\densityw}(191.2,-30.3)(193.4,-27.0)(195.1,-23.3)
\bezier{\densityw}(195.1,-23.3)(196.7,-19.7)(197.9,-15.8)
\bezier{\densityw}(197.9,-15.8)(199.0,-12.0)(199.6,-8.0)
\bezier{\densityw}(199.6,-8.0)(200.2,-4.0)(200.2,-0.0)
\bezier{\densityw}(200.2,-0.0)(200.2, 4.0)(199.6, 8.0)
\bezier{\densityw}(199.6, 8.0)(199.0,11.9)(197.9,15.8)
\bezier{\densityw}(197.9,15.8)(196.7,19.6)(195.1,23.3)
\bezier{\densityw}(195.1,23.3)(193.4,26.9)(191.2,30.3)
\bezier{\densityw}(191.2,30.3)(189.1,33.7)(186.5,36.7)
\bezier{\densityw}(186.5,36.7)(183.8,39.7)(180.8,42.4)
\bezier{\densityw}(180.8,42.4)(177.8,45.0)(174.4,47.2)
\bezier{\densityw}(174.4,47.2)(171.0,49.3)(167.4,51.0)
\bezier{\densityw}(167.4,51.0)(163.7,52.7)(159.9,53.8)
\bezier{\densityw}(159.9,53.8)(156.0,54.9)(152.1,55.5)
\bezier{\densityw}(152.1,55.5)(148.1,56.1)(144.1,56.1)
\put(200.2,-0.0){\line(1,0){88.0}}
\bezier{\densityw}(144.1,56.1)(140.1,56.1)(136.1,55.5)
\bezier{\densityw}(136.1,55.5)(132.1,54.9)(128.3,53.8)
\bezier{\densityw}(128.3,53.8)(124.4,52.7)(120.8,51.0)
\bezier{\densityw}(120.8,51.0)(117.1,49.3)(113.8,47.2)
\bezier{\densityw}(113.8,47.2)(110.4,45.0)(107.4,42.4)
\bezier{\densityw}(107.4,42.4)(104.3,39.7)(101.7,36.7)
\bezier{\densityw}(101.7,36.7)(99.1,33.7)(96.9,30.3)
\bezier{\densityw}(96.9,30.3)(94.7,26.9)(93.1,23.3)
\bezier{\densityw}(93.1,23.3)(91.4,19.6)(90.3,15.8)
\bezier{\densityw}(90.3,15.8)(89.2,11.9)(88.6, 8.0)
\bezier{\densityw}(88.6, 8.0)(88.0, 4.0)(88.0,-0.0)
\put(144.1,56.1){\line(0,-1){112.1}}
\end{picture}
} }
\newcommand{\subc}{
\setlength{\unitlength}{0.069mm}
\raisebox{-56.1\unitlength}{
\begin{picture}(288.2,112.2)( 0.0,-56.1)
\put( 0.0, 0.0){\line(1,0){88.0}}
\bezier{\densityw}(88.0, 0.0)(88.0,-4.0)(88.6,-8.0)
\bezier{\densityw}(88.6,-8.0)(89.1,-11.9)(90.3,-15.8)
\bezier{\densityw}(90.3,-15.8)(91.4,-19.6)(93.1,-23.3)
\bezier{\densityw}(93.1,-23.3)(94.7,-26.9)(96.9,-30.3)
\bezier{\densityw}(96.9,-30.3)(99.1,-33.7)(101.7,-36.7)
\bezier{\densityw}(101.7,-36.7)(104.3,-39.7)(107.4,-42.4)
\bezier{\densityw}(107.4,-42.4)(110.4,-45.0)(113.8,-47.2)
\bezier{\densityw}(113.8,-47.2)(117.1,-49.3)(120.8,-51.0)
\bezier{\densityw}(120.8,-51.0)(124.4,-52.7)(128.3,-53.8)
\bezier{\densityw}(128.3,-53.8)(132.1,-54.9)(136.1,-55.5)
\bezier{\densityw}(136.1,-55.5)(140.1,-56.1)(144.1,-56.1)
\bezier{\densityw}(144.1,-56.1)(144.5,-56.1)(145.0,-56.1)
\bezier{\densityw}(152.8,-55.4)(153.3,-55.3)(153.7,-55.2)
\bezier{\densityw}(161.4,-53.3)(161.8,-53.2)(162.2,-53.0)
\bezier{\densityw}(169.5,-50.0)(169.9,-49.8)(170.3,-49.6)
\bezier{\densityw}(177.0,-45.4)(177.4,-45.1)(177.7,-44.8)
\bezier{\densityw}(183.7,-39.6)(184.0,-39.3)(184.3,-39.0)
\bezier{\densityw}(189.4,-33.0)(189.7,-32.6)(190.0,-32.2)
\bezier{\densityw}(194.0,-25.4)(194.2,-25.1)(194.4,-24.7)
\bezier{\densityw}(197.4,-17.3)(197.5,-16.9)(197.7,-16.5)
\bezier{\densityw}(199.5,-8.8)(199.5,-8.3)(199.6,-7.9)
\bezier{\densityw}(200.2, 0.0)(200.2, 4.0)(199.6, 8.0)
\bezier{\densityw}(199.6, 8.0)(199.0,12.0)(197.9,15.8)
\bezier{\densityw}(197.9,15.8)(196.7,19.7)(195.1,23.3)
\bezier{\densityw}(195.1,23.3)(193.4,27.0)(191.2,30.3)
\bezier{\densityw}(191.2,30.3)(189.1,33.7)(186.5,36.7)
\bezier{\densityw}(186.5,36.7)(183.8,39.8)(180.8,42.4)
\bezier{\densityw}(180.8,42.4)(177.8,45.0)(174.4,47.2)
\bezier{\densityw}(174.4,47.2)(171.0,49.3)(167.4,51.0)
\bezier{\densityw}(167.4,51.0)(163.7,52.7)(159.9,53.8)
\bezier{\densityw}(159.9,53.8)(156.0,54.9)(152.1,55.5)
\bezier{\densityw}(152.1,55.5)(148.1,56.1)(144.1,56.1)
\put(200.2, 0.0){\line(1,0){88.0}}
\bezier{\densityw}(144.1,56.1)(140.1,56.1)(136.1,55.5)
\bezier{\densityw}(136.1,55.5)(132.1,54.9)(128.3,53.8)
\bezier{\densityw}(128.3,53.8)(124.4,52.7)(120.8,51.0)
\bezier{\densityw}(120.8,51.0)(117.1,49.3)(113.8,47.2)
\bezier{\densityw}(113.8,47.2)(110.4,45.0)(107.4,42.4)
\bezier{\densityw}(107.4,42.4)(104.3,39.8)(101.7,36.7)
\bezier{\densityw}(101.7,36.7)(99.1,33.7)(96.9,30.3)
\bezier{\densityw}(96.9,30.3)(94.7,27.0)(93.1,23.3)
\bezier{\densityw}(93.1,23.3)(91.4,19.7)(90.3,15.8)
\bezier{\densityw}(90.3,15.8)(89.2,12.0)(88.6, 8.0)
\bezier{\densityw}(88.6, 8.0)(88.0, 4.0)(88.0, 0.0)
\put(144.1,56.1){\line(0,-1){112.2}}
\end{picture}
} }
\newcommand{\subd}{
\setlength{\unitlength}{0.069mm}
\raisebox{-56.1\unitlength}{
\begin{picture}(288.1,112.1)( 0.0,-56.1)
\put( 0.0, 0.0){\line(1,0){88.0}}
\bezier{\densityw}(88.0, 0.0)(88.0,-4.0)(88.6,-8.0)
\bezier{\densityw}(88.6,-8.0)(89.1,-11.9)(90.3,-15.8)
\bezier{\densityw}(90.3,-15.8)(91.4,-19.6)(93.1,-23.3)
\bezier{\densityw}(93.1,-23.3)(94.7,-26.9)(96.9,-30.3)
\bezier{\densityw}(96.9,-30.3)(99.1,-33.7)(101.7,-36.7)
\bezier{\densityw}(101.7,-36.7)(104.3,-39.7)(107.4,-42.4)
\bezier{\densityw}(107.4,-42.4)(110.4,-45.0)(113.8,-47.2)
\bezier{\densityw}(113.8,-47.2)(117.1,-49.3)(120.8,-51.0)
\bezier{\densityw}(120.8,-51.0)(124.4,-52.7)(128.3,-53.8)
\bezier{\densityw}(128.3,-53.8)(132.1,-54.9)(136.1,-55.5)
\bezier{\densityw}(136.1,-55.5)(140.1,-56.1)(144.1,-56.1)
\bezier{\densityw}(144.1,-56.1)(148.1,-56.1)(152.0,-55.5)
\bezier{\densityw}(152.0,-55.5)(156.0,-54.9)(159.9,-53.8)
\bezier{\densityw}(159.9,-53.8)(163.7,-52.7)(167.4,-51.0)
\bezier{\densityw}(167.4,-51.0)(171.0,-49.3)(174.4,-47.2)
\bezier{\densityw}(174.4,-47.2)(177.8,-45.0)(180.8,-42.4)
\bezier{\densityw}(180.8,-42.4)(183.8,-39.7)(186.4,-36.7)
\bezier{\densityw}(186.4,-36.7)(189.1,-33.7)(191.2,-30.3)
\bezier{\densityw}(191.2,-30.3)(193.4,-26.9)(195.1,-23.3)
\bezier{\densityw}(195.1,-23.3)(196.7,-19.6)(197.9,-15.8)
\bezier{\densityw}(197.9,-15.8)(199.0,-11.9)(199.6,-8.0)
\bezier{\densityw}(199.6,-8.0)(200.1,-4.0)(200.1, 0.0)
\bezier{\densityw}(200.1, 0.0)(200.1, 4.0)(199.6, 8.0)
\bezier{\densityw}(199.6, 8.0)(199.0,11.9)(197.9,15.8)
\bezier{\densityw}(197.9,15.8)(196.7,19.6)(195.1,23.3)
\bezier{\densityw}(195.1,23.3)(193.4,26.9)(191.2,30.3)
\bezier{\densityw}(191.2,30.3)(189.1,33.7)(186.4,36.7)
\bezier{\densityw}(186.4,36.7)(183.8,39.7)(180.8,42.4)
\bezier{\densityw}(180.8,42.4)(177.8,45.0)(174.4,47.2)
\bezier{\densityw}(174.4,47.2)(171.0,49.3)(167.4,51.0)
\bezier{\densityw}(167.4,51.0)(163.7,52.7)(159.9,53.8)
\bezier{\densityw}(159.9,53.8)(156.0,54.9)(152.0,55.5)
\bezier{\densityw}(152.0,55.5)(148.1,56.1)(144.1,56.1)
\put(200.1, 0.0){\line(1,0){88.0}}
\bezier{\densityw}(144.1,56.1)(143.6,56.1)(143.2,56.1)
\bezier{\densityw}(135.3,55.4)(134.9,55.3)(134.4,55.2)
\bezier{\densityw}(126.7,53.3)(126.3,53.2)(125.9,53.0)
\bezier{\densityw}(118.6,50.0)(118.2,49.8)(117.8,49.6)
\bezier{\densityw}(111.1,45.4)(110.8,45.1)(110.4,44.8)
\bezier{\densityw}(104.4,39.6)(104.1,39.3)(103.8,39.0)
\bezier{\densityw}(98.7,33.0)(98.4,32.6)(98.2,32.2)
\bezier{\densityw}(94.1,25.4)(93.9,25.1)(93.7,24.7)
\bezier{\densityw}(90.7,17.3)(90.6,16.9)(90.5,16.5)
\bezier{\densityw}(88.7, 8.8)(88.6, 8.3)(88.5, 7.9)
\put(144.1,56.1){\line(0,-1){112.1}}
\end{picture}
} }
\newcommand{\sube}{
\setlength{\unitlength}{0.069mm}
\raisebox{-56.1\unitlength}{
\begin{picture}(288.1,112.2)( 0.0,-56.1)
\put( 0.0, 0.0){\line(1,0){88.0}}
\bezier{\densityw}(88.0, 0.0)(88.0,-4.0)(88.6,-8.0)
\bezier{\densityw}(88.6,-8.0)(89.1,-11.9)(90.3,-15.8)
\bezier{\densityw}(90.3,-15.8)(91.4,-19.6)(93.1,-23.3)
\bezier{\densityw}(93.1,-23.3)(94.7,-26.9)(96.9,-30.3)
\bezier{\densityw}(96.9,-30.3)(99.1,-33.7)(101.7,-36.7)
\bezier{\densityw}(101.7,-36.7)(104.3,-39.7)(107.4,-42.4)
\bezier{\densityw}(107.4,-42.4)(110.4,-45.0)(113.8,-47.2)
\bezier{\densityw}(113.8,-47.2)(117.1,-49.3)(120.8,-51.0)
\bezier{\densityw}(120.8,-51.0)(124.4,-52.7)(128.3,-53.8)
\bezier{\densityw}(128.3,-53.8)(132.1,-54.9)(136.1,-55.5)
\bezier{\densityw}(136.1,-55.5)(140.1,-56.1)(144.1,-56.1)
\bezier{\densityw}(144.1,-56.1)(148.1,-56.1)(152.0,-55.5)
\bezier{\densityw}(152.0,-55.5)(156.0,-54.9)(159.9,-53.8)
\bezier{\densityw}(159.9,-53.8)(163.7,-52.7)(167.4,-51.0)
\bezier{\densityw}(167.4,-51.0)(171.0,-49.3)(174.4,-47.2)
\bezier{\densityw}(174.4,-47.2)(177.8,-45.0)(180.8,-42.4)
\bezier{\densityw}(180.8,-42.4)(183.8,-39.7)(186.4,-36.7)
\bezier{\densityw}(186.4,-36.7)(189.1,-33.7)(191.2,-30.3)
\bezier{\densityw}(191.2,-30.3)(193.4,-26.9)(195.1,-23.3)
\bezier{\densityw}(195.1,-23.3)(196.7,-19.6)(197.9,-15.8)
\bezier{\densityw}(197.9,-15.8)(199.0,-11.9)(199.6,-8.0)
\bezier{\densityw}(199.6,-8.0)(200.1,-4.0)(200.1, 0.0)
\bezier{\densityw}(200.1, 0.0)(200.1, 0.4)(200.1, 0.9)
\bezier{\densityw}(199.4, 8.8)(199.4, 9.2)(199.3, 9.6)
\bezier{\densityw}(197.4,17.3)(197.3,17.7)(197.1,18.2)
\bezier{\densityw}(194.0,25.5)(193.8,25.9)(193.6,26.2)
\bezier{\densityw}(189.4,33.0)(189.2,33.3)(188.9,33.7)
\bezier{\densityw}(183.7,39.7)(183.4,40.0)(183.1,40.3)
\bezier{\densityw}(177.0,45.4)(176.7,45.6)(176.3,45.9)
\bezier{\densityw}(169.5,50.0)(169.1,50.2)(168.7,50.4)
\bezier{\densityw}(161.4,53.3)(161.0,53.5)(160.5,53.6)
\bezier{\densityw}(152.8,55.4)(152.4,55.5)(152.0,55.5)
\put(200.1, 0.0){\line(1,0){88.0}}
\bezier{\densityw}(144.1,56.1)(140.0,56.1)(136.1,55.5)
\bezier{\densityw}(136.1,55.5)(132.1,54.9)(128.3,53.8)
\bezier{\densityw}(128.3,53.8)(124.4,52.7)(120.8,51.0)
\bezier{\densityw}(120.8,51.0)(117.1,49.3)(113.7,47.2)
\bezier{\densityw}(113.7,47.2)(110.4,45.0)(107.3,42.4)
\bezier{\densityw}(107.3,42.4)(104.3,39.8)(101.7,36.7)
\bezier{\densityw}(101.7,36.7)(99.1,33.7)(96.9,30.3)
\bezier{\densityw}(96.9,30.3)(94.7,27.0)(93.1,23.3)
\bezier{\densityw}(93.1,23.3)(91.4,19.7)(90.3,15.8)
\bezier{\densityw}(90.3,15.8)(89.1,12.0)(88.6, 8.0)
\bezier{\densityw}(88.6, 8.0)(88.0, 4.0)(88.0, 0.0)
\put(144.1,56.1){\line(0,-1){112.2}}
\end{picture}
} }
\newcommand{\subf}{
\setlength{\unitlength}{0.069mm}
\raisebox{-56.1\unitlength}{
\begin{picture}(288.1,112.1)( 0.0,-56.1)
\put( 0.0, 0.0){\line(1,0){88.0}}
\bezier{\densityw}(88.0, 0.0)(88.0,-4.0)(88.6,-8.0)
\bezier{\densityw}(88.6,-8.0)(89.1,-11.9)(90.3,-15.8)
\bezier{\densityw}(90.3,-15.8)(91.4,-19.6)(93.1,-23.3)
\bezier{\densityw}(93.1,-23.3)(94.7,-26.9)(96.9,-30.3)
\bezier{\densityw}(96.9,-30.3)(99.1,-33.7)(101.7,-36.7)
\bezier{\densityw}(101.7,-36.7)(104.3,-39.7)(107.4,-42.4)
\bezier{\densityw}(107.4,-42.4)(110.4,-45.0)(113.8,-47.2)
\bezier{\densityw}(113.8,-47.2)(117.1,-49.3)(120.8,-51.0)
\bezier{\densityw}(120.8,-51.0)(124.4,-52.7)(128.3,-53.8)
\bezier{\densityw}(128.3,-53.8)(132.1,-54.9)(136.1,-55.5)
\bezier{\densityw}(136.1,-55.5)(140.1,-56.1)(144.1,-56.1)
\bezier{\densityw}(144.1,-56.1)(148.1,-56.1)(152.0,-55.5)
\bezier{\densityw}(152.0,-55.5)(156.0,-54.9)(159.9,-53.8)
\bezier{\densityw}(159.9,-53.8)(163.7,-52.7)(167.4,-51.0)
\bezier{\densityw}(167.4,-51.0)(171.0,-49.3)(174.4,-47.2)
\bezier{\densityw}(174.4,-47.2)(177.8,-45.0)(180.8,-42.4)
\bezier{\densityw}(180.8,-42.4)(183.8,-39.7)(186.4,-36.7)
\bezier{\densityw}(186.4,-36.7)(189.1,-33.7)(191.2,-30.3)
\bezier{\densityw}(191.2,-30.3)(193.4,-26.9)(195.1,-23.3)
\bezier{\densityw}(195.1,-23.3)(196.7,-19.6)(197.9,-15.8)
\bezier{\densityw}(197.9,-15.8)(199.0,-11.9)(199.6,-8.0)
\bezier{\densityw}(199.6,-8.0)(200.1,-4.0)(200.1, 0.0)
\bezier{\densityw}(200.1, 0.0)(200.1, 4.0)(199.6, 8.0)
\bezier{\densityw}(199.6, 8.0)(199.0,11.9)(197.9,15.8)
\bezier{\densityw}(197.9,15.8)(196.7,19.6)(195.1,23.3)
\bezier{\densityw}(195.1,23.3)(193.4,26.9)(191.2,30.3)
\bezier{\densityw}(191.2,30.3)(189.1,33.7)(186.4,36.7)
\bezier{\densityw}(186.4,36.7)(183.8,39.7)(180.8,42.4)
\bezier{\densityw}(180.8,42.4)(177.8,45.0)(174.4,47.2)
\bezier{\densityw}(174.4,47.2)(171.0,49.3)(167.4,51.0)
\bezier{\densityw}(167.4,51.0)(163.7,52.7)(159.9,53.8)
\bezier{\densityw}(159.9,53.8)(156.0,54.9)(152.0,55.5)
\bezier{\densityw}(152.0,55.5)(148.1,56.1)(144.1,56.1)
\put(200.1, 0.0){\line(1,0){88.0}}
\bezier{\densityw}(144.1,56.1)(140.1,56.1)(136.1,55.5)
\bezier{\densityw}(136.1,55.5)(132.1,54.9)(128.3,53.8)
\bezier{\densityw}(128.3,53.8)(124.4,52.7)(120.8,51.0)
\bezier{\densityw}(120.8,51.0)(117.1,49.3)(113.8,47.2)
\bezier{\densityw}(113.8,47.2)(110.4,45.0)(107.4,42.4)
\bezier{\densityw}(107.4,42.4)(104.3,39.7)(101.7,36.7)
\bezier{\densityw}(101.7,36.7)(99.1,33.7)(96.9,30.3)
\bezier{\densityw}(96.9,30.3)(94.7,26.9)(93.1,23.3)
\bezier{\densityw}(93.1,23.3)(91.4,19.6)(90.3,15.8)
\bezier{\densityw}(90.3,15.8)(89.1,11.9)(88.6, 8.0)
\bezier{\densityw}(88.6, 8.0)(88.0, 4.0)(88.0, 0.0)
\multiput(144.1,56.1)(-0.00,-8.63){13}{\line(0,-1){ 0.9}}
\end{picture}
} }
\newcommand{\subg}{
\setlength{\unitlength}{0.069mm}
\raisebox{-56.1\unitlength}{
\begin{picture}(288.2,112.2)( 0.0,-56.1)
\put( 0.0, 0.0){\line(1,0){88.0}}
\bezier{\densityw}(88.0, 0.0)(88.0,-0.4)(88.0,-0.9)
\bezier{\densityw}(88.7,-8.8)(88.8,-9.2)(88.8,-9.6)
\bezier{\densityw}(90.7,-17.3)(90.9,-17.7)(91.0,-18.2)
\bezier{\densityw}(94.1,-25.5)(94.3,-25.9)(94.5,-26.2)
\bezier{\densityw}(98.7,-33.0)(99.0,-33.3)(99.2,-33.7)
\bezier{\densityw}(104.4,-39.7)(104.7,-40.0)(105.1,-40.3)
\bezier{\densityw}(111.1,-45.4)(111.5,-45.6)(111.8,-45.9)
\bezier{\densityw}(118.6,-50.0)(119.0,-50.2)(119.4,-50.4)
\bezier{\densityw}(126.8,-53.3)(127.2,-53.5)(127.6,-53.6)
\bezier{\densityw}(135.3,-55.4)(135.7,-55.5)(136.2,-55.5)
\bezier{\densityw}(144.1,-56.1)(148.1,-56.1)(152.1,-55.5)
\bezier{\densityw}(152.1,-55.5)(156.0,-54.9)(159.9,-53.8)
\bezier{\densityw}(159.9,-53.8)(163.7,-52.7)(167.4,-51.0)
\bezier{\densityw}(167.4,-51.0)(171.0,-49.3)(174.4,-47.2)
\bezier{\densityw}(174.4,-47.2)(177.8,-45.0)(180.8,-42.4)
\bezier{\densityw}(180.8,-42.4)(183.8,-39.8)(186.5,-36.7)
\bezier{\densityw}(186.5,-36.7)(189.1,-33.7)(191.2,-30.3)
\bezier{\densityw}(191.2,-30.3)(193.4,-27.0)(195.1,-23.3)
\bezier{\densityw}(195.1,-23.3)(196.7,-19.7)(197.9,-15.8)
\bezier{\densityw}(197.9,-15.8)(199.0,-12.0)(199.6,-8.0)
\bezier{\densityw}(199.6,-8.0)(200.2,-4.0)(200.2,-0.0)
\bezier{\densityw}(200.2,-0.0)(200.2, 0.4)(200.1, 0.9)
\bezier{\densityw}(199.5, 8.8)(199.4, 9.2)(199.3, 9.6)
\bezier{\densityw}(197.4,17.3)(197.3,17.7)(197.1,18.2)
\bezier{\densityw}(194.0,25.4)(193.8,25.8)(193.6,26.2)
\bezier{\densityw}(189.4,33.0)(189.2,33.3)(188.9,33.7)
\bezier{\densityw}(183.7,39.6)(183.4,40.0)(183.1,40.3)
\bezier{\densityw}(177.0,45.4)(176.7,45.6)(176.3,45.9)
\bezier{\densityw}(169.5,50.0)(169.1,50.2)(168.7,50.4)
\bezier{\densityw}(161.4,53.3)(161.0,53.5)(160.6,53.6)
\bezier{\densityw}(152.8,55.4)(152.4,55.4)(152.0,55.5)
\put(200.2,-0.0){\line(1,0){88.0}}
\bezier{\densityw}(144.1,56.1)(140.1,56.1)(136.1,55.5)
\bezier{\densityw}(136.1,55.5)(132.1,54.9)(128.3,53.8)
\bezier{\densityw}(128.3,53.8)(124.4,52.7)(120.8,51.0)
\bezier{\densityw}(120.8,51.0)(117.1,49.3)(113.8,47.2)
\bezier{\densityw}(113.8,47.2)(110.4,45.0)(107.4,42.4)
\bezier{\densityw}(107.4,42.4)(104.3,39.7)(101.7,36.7)
\bezier{\densityw}(101.7,36.7)(99.1,33.7)(96.9,30.3)
\bezier{\densityw}(96.9,30.3)(94.7,26.9)(93.1,23.3)
\bezier{\densityw}(93.1,23.3)(91.4,19.6)(90.3,15.8)
\bezier{\densityw}(90.3,15.8)(89.1,11.9)(88.6, 8.0)
\bezier{\densityw}(88.6, 8.0)(88.0, 4.0)(88.0, 0.0)
\put(144.1,56.1){\line(0,-1){112.2}}
\end{picture}
} }
\newcommand{\subh}{
\setlength{\unitlength}{0.069mm}
\raisebox{-56.1\unitlength}{
\begin{picture}(288.2,112.2)( 0.0,-56.1)
\put( 0.0, 0.0){\line(1,0){88.0}}
\bezier{\densityw}(88.0, 0.0)(88.0,-4.0)(88.6,-8.0)
\bezier{\densityw}(88.6,-8.0)(89.1,-11.9)(90.3,-15.8)
\bezier{\densityw}(90.3,-15.8)(91.4,-19.6)(93.1,-23.3)
\bezier{\densityw}(93.1,-23.3)(94.7,-26.9)(96.9,-30.3)
\bezier{\densityw}(96.9,-30.3)(99.1,-33.7)(101.7,-36.7)
\bezier{\densityw}(101.7,-36.7)(104.3,-39.7)(107.4,-42.4)
\bezier{\densityw}(107.4,-42.4)(110.4,-45.0)(113.8,-47.2)
\bezier{\densityw}(113.8,-47.2)(117.1,-49.3)(120.8,-51.0)
\bezier{\densityw}(120.8,-51.0)(124.4,-52.7)(128.3,-53.8)
\bezier{\densityw}(128.3,-53.8)(132.1,-54.9)(136.1,-55.5)
\bezier{\densityw}(136.1,-55.5)(140.1,-56.1)(144.1,-56.1)
\bezier{\densityw}(144.1,-56.1)(144.5,-56.1)(145.0,-56.1)
\bezier{\densityw}(152.8,-55.4)(153.3,-55.3)(153.7,-55.2)
\bezier{\densityw}(161.4,-53.3)(161.8,-53.2)(162.2,-53.0)
\bezier{\densityw}(169.5,-50.0)(169.9,-49.8)(170.3,-49.6)
\bezier{\densityw}(177.0,-45.4)(177.4,-45.1)(177.7,-44.8)
\bezier{\densityw}(183.7,-39.6)(184.0,-39.3)(184.3,-39.0)
\bezier{\densityw}(189.4,-33.0)(189.7,-32.6)(190.0,-32.2)
\bezier{\densityw}(194.0,-25.4)(194.2,-25.1)(194.4,-24.7)
\bezier{\densityw}(197.4,-17.3)(197.5,-16.9)(197.7,-16.5)
\bezier{\densityw}(199.5,-8.8)(199.5,-8.3)(199.6,-7.9)
\bezier{\densityw}(200.2, 0.0)(200.2, 4.0)(199.6, 8.0)
\bezier{\densityw}(199.6, 8.0)(199.0,12.0)(197.9,15.8)
\bezier{\densityw}(197.9,15.8)(196.7,19.7)(195.1,23.3)
\bezier{\densityw}(195.1,23.3)(193.4,27.0)(191.2,30.3)
\bezier{\densityw}(191.2,30.3)(189.1,33.7)(186.5,36.7)
\bezier{\densityw}(186.5,36.7)(183.8,39.8)(180.8,42.4)
\bezier{\densityw}(180.8,42.4)(177.8,45.0)(174.4,47.2)
\bezier{\densityw}(174.4,47.2)(171.0,49.3)(167.4,51.0)
\bezier{\densityw}(167.4,51.0)(163.7,52.7)(159.9,53.8)
\bezier{\densityw}(159.9,53.8)(156.0,54.9)(152.1,55.5)
\bezier{\densityw}(152.1,55.5)(148.1,56.1)(144.1,56.1)
\put(200.2, 0.0){\line(1,0){88.0}}
\bezier{\densityw}(144.1,56.1)(143.6,56.1)(143.2,56.1)
\bezier{\densityw}(135.3,55.4)(134.9,55.3)(134.4,55.2)
\bezier{\densityw}(126.8,53.3)(126.3,53.2)(125.9,53.1)
\bezier{\densityw}(118.6,50.0)(118.2,49.8)(117.8,49.6)
\bezier{\densityw}(111.1,45.4)(110.8,45.1)(110.4,44.8)
\bezier{\densityw}(104.4,39.7)(104.1,39.3)(103.8,39.0)
\bezier{\densityw}(98.7,33.0)(98.5,32.6)(98.2,32.2)
\bezier{\densityw}(94.1,25.5)(93.9,25.1)(93.7,24.7)
\bezier{\densityw}(90.7,17.3)(90.6,16.9)(90.5,16.5)
\bezier{\densityw}(88.7, 8.8)(88.6, 8.3)(88.6, 7.9)
\put(144.1,56.1){\line(0,-1){112.2}}
\end{picture}
} }
\newcommand{\subi}{
\setlength{\unitlength}{0.069mm}
\raisebox{-56.1\unitlength}{
\begin{picture}(288.2,112.2)( 0.0,-56.1)
\put( 0.0, 0.0){\line(1,0){88.0}}
\bezier{\densityw}(88.0, 0.0)(88.0,-0.4)(88.0,-0.9)
\bezier{\densityw}(88.7,-8.8)(88.8,-9.2)(88.8,-9.6)
\bezier{\densityw}(90.7,-17.3)(90.9,-17.7)(91.0,-18.2)
\bezier{\densityw}(94.1,-25.5)(94.3,-25.9)(94.5,-26.2)
\bezier{\densityw}(98.7,-33.0)(99.0,-33.3)(99.2,-33.7)
\bezier{\densityw}(104.4,-39.7)(104.7,-40.0)(105.1,-40.3)
\bezier{\densityw}(111.1,-45.4)(111.5,-45.6)(111.8,-45.9)
\bezier{\densityw}(118.6,-50.0)(119.0,-50.2)(119.4,-50.4)
\bezier{\densityw}(126.8,-53.3)(127.2,-53.5)(127.6,-53.6)
\bezier{\densityw}(135.3,-55.4)(135.7,-55.5)(136.2,-55.5)
\bezier{\densityw}(144.1,-56.1)(144.5,-56.1)(145.0,-56.1)
\bezier{\densityw}(152.9,-55.4)(153.3,-55.3)(153.7,-55.2)
\bezier{\densityw}(161.4,-53.3)(161.8,-53.2)(162.2,-53.1)
\bezier{\densityw}(169.5,-50.0)(169.9,-49.8)(170.3,-49.6)
\bezier{\densityw}(177.0,-45.4)(177.4,-45.1)(177.8,-44.8)
\bezier{\densityw}(183.7,-39.7)(184.0,-39.3)(184.4,-39.0)
\bezier{\densityw}(189.5,-33.0)(189.7,-32.6)(190.0,-32.2)
\bezier{\densityw}(194.0,-25.5)(194.2,-25.1)(194.4,-24.7)
\bezier{\densityw}(197.4,-17.3)(197.6,-16.9)(197.7,-16.5)
\bezier{\densityw}(199.5,-8.8)(199.5,-8.3)(199.6,-7.9)
\bezier{\densityw}(200.2, 0.0)(200.2, 4.0)(199.6, 8.0)
\bezier{\densityw}(199.6, 8.0)(199.0,11.9)(197.9,15.8)
\bezier{\densityw}(197.9,15.8)(196.8,19.6)(195.1,23.3)
\bezier{\densityw}(195.1,23.3)(193.4,26.9)(191.3,30.3)
\bezier{\densityw}(191.3,30.3)(189.1,33.7)(186.5,36.7)
\bezier{\densityw}(186.5,36.7)(183.8,39.7)(180.8,42.4)
\bezier{\densityw}(180.8,42.4)(177.8,45.0)(174.4,47.2)
\bezier{\densityw}(174.4,47.2)(171.0,49.3)(167.4,51.0)
\bezier{\densityw}(167.4,51.0)(163.7,52.7)(159.9,53.8)
\bezier{\densityw}(159.9,53.8)(156.0,54.9)(152.1,55.5)
\bezier{\densityw}(152.1,55.5)(148.1,56.1)(144.1,56.1)
\put(200.2, 0.0){\line(1,0){88.0}}
\bezier{\densityw}(144.1,56.1)(140.1,56.1)(136.1,55.5)
\bezier{\densityw}(136.1,55.5)(132.1,54.9)(128.3,53.8)
\bezier{\densityw}(128.3,53.8)(124.4,52.7)(120.8,51.0)
\bezier{\densityw}(120.8,51.0)(117.2,49.3)(113.8,47.2)
\bezier{\densityw}(113.8,47.2)(110.4,45.0)(107.4,42.4)
\bezier{\densityw}(107.4,42.4)(104.3,39.7)(101.7,36.7)
\bezier{\densityw}(101.7,36.7)(99.1,33.7)(96.9,30.3)
\bezier{\densityw}(96.9,30.3)(94.8,26.9)(93.1,23.3)
\bezier{\densityw}(93.1,23.3)(91.4,19.6)(90.3,15.8)
\bezier{\densityw}(90.3,15.8)(89.2,11.9)(88.6, 8.0)
\bezier{\densityw}(88.6, 8.0)(88.0, 4.0)(88.0, 0.0)
\multiput(144.1,56.1)(-0.00,-8.63){13}{\line(0,-1){ 0.9}}
\end{picture}
} }
\newcommand{\subj}{
\setlength{\unitlength}{0.069mm}
\raisebox{-56.1\unitlength}{
\begin{picture}(288.1,112.2)( 0.0,-56.1)
\put( 0.0, 0.0){\line(1,0){88.0}}
\bezier{\densityw}(88.0, 0.0)(88.0,-4.0)(88.6,-8.0)
\bezier{\densityw}(88.6,-8.0)(89.1,-11.9)(90.3,-15.8)
\bezier{\densityw}(90.3,-15.8)(91.4,-19.6)(93.1,-23.3)
\bezier{\densityw}(93.1,-23.3)(94.7,-26.9)(96.9,-30.3)
\bezier{\densityw}(96.9,-30.3)(99.1,-33.7)(101.7,-36.7)
\bezier{\densityw}(101.7,-36.7)(104.3,-39.7)(107.4,-42.4)
\bezier{\densityw}(107.4,-42.4)(110.4,-45.0)(113.8,-47.2)
\bezier{\densityw}(113.8,-47.2)(117.1,-49.3)(120.8,-51.0)
\bezier{\densityw}(120.8,-51.0)(124.4,-52.7)(128.3,-53.8)
\bezier{\densityw}(128.3,-53.8)(132.1,-54.9)(136.1,-55.5)
\bezier{\densityw}(136.1,-55.5)(140.1,-56.1)(144.1,-56.1)
\bezier{\densityw}(144.1,-56.1)(148.1,-56.1)(152.0,-55.5)
\bezier{\densityw}(152.0,-55.5)(156.0,-54.9)(159.9,-53.8)
\bezier{\densityw}(159.9,-53.8)(163.7,-52.7)(167.4,-51.0)
\bezier{\densityw}(167.4,-51.0)(171.0,-49.3)(174.4,-47.2)
\bezier{\densityw}(174.4,-47.2)(177.8,-45.0)(180.8,-42.4)
\bezier{\densityw}(180.8,-42.4)(183.8,-39.7)(186.4,-36.7)
\bezier{\densityw}(186.4,-36.7)(189.1,-33.7)(191.2,-30.3)
\bezier{\densityw}(191.2,-30.3)(193.4,-26.9)(195.1,-23.3)
\bezier{\densityw}(195.1,-23.3)(196.7,-19.6)(197.9,-15.8)
\bezier{\densityw}(197.9,-15.8)(199.0,-11.9)(199.6,-8.0)
\bezier{\densityw}(199.6,-8.0)(200.1,-4.0)(200.1, 0.0)
\bezier{\densityw}(200.1, 0.0)(200.1, 0.4)(200.1, 0.9)
\bezier{\densityw}(199.4, 8.8)(199.4, 9.2)(199.3, 9.6)
\bezier{\densityw}(197.4,17.3)(197.3,17.7)(197.1,18.2)
\bezier{\densityw}(194.0,25.5)(193.8,25.9)(193.6,26.2)
\bezier{\densityw}(189.4,33.0)(189.2,33.3)(188.9,33.7)
\bezier{\densityw}(183.7,39.7)(183.4,40.0)(183.1,40.3)
\bezier{\densityw}(177.0,45.4)(176.7,45.6)(176.3,45.9)
\bezier{\densityw}(169.5,50.0)(169.1,50.2)(168.7,50.4)
\bezier{\densityw}(161.4,53.3)(161.0,53.5)(160.5,53.6)
\bezier{\densityw}(152.8,55.4)(152.4,55.5)(152.0,55.5)
\put(200.1, 0.0){\line(1,0){88.0}}
\bezier{\densityw}(144.1,56.1)(143.6,56.1)(143.2,56.1)
\bezier{\densityw}(135.3,55.4)(134.9,55.3)(134.4,55.2)
\bezier{\densityw}(126.7,53.3)(126.3,53.2)(125.9,53.1)
\bezier{\densityw}(118.6,50.0)(118.2,49.8)(117.8,49.6)
\bezier{\densityw}(111.1,45.4)(110.7,45.1)(110.4,44.8)
\bezier{\densityw}(104.4,39.7)(104.1,39.3)(103.8,39.0)
\bezier{\densityw}(98.7,33.0)(98.4,32.6)(98.2,32.2)
\bezier{\densityw}(94.1,25.5)(93.9,25.1)(93.7,24.7)
\bezier{\densityw}(90.7,17.3)(90.6,16.9)(90.5,16.5)
\bezier{\densityw}(88.7, 8.8)(88.6, 8.3)(88.5, 7.9)
\multiput(144.1,56.1)(0.00,-8.63){13}{\line(0,-1){ 0.9}}
\end{picture}
} }

\begin{figure}[htb]
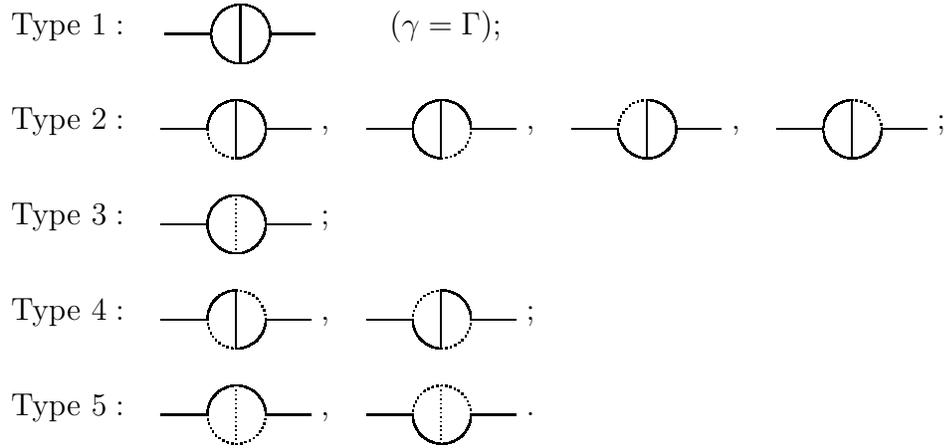

  \[
  \begin{array}{rcccc}
\mbox{ Type 1}:&\suba & (\gamma=\Gamma);&      &      \\[4ex]
\mbox{ Type 2}:&\subb,&\subc,           &\subd,&\sube;\\[4ex]
\mbox{ Type 3}:&\subf;&                 &      &      \\[4ex]
\mbox{ Type 4}:&\subg,&\subh;           &      &      \\[4ex]
\mbox{ Type 5}:&\subi,&\subj.           &      &
  \end{array}
  \]
  \caption{ The subgraphs $\gamma$ contributing to the large $k^2$ expansion }
  \label{subgraphs}
\end{figure}
The reduced graphs $\Gamma/\gamma$ correspond to the dotted lines and
can be obtained by shrinking all solid lines to a point.  In such a
way, we obtain that for the second and third type (see Fig.2)
$J_{\Gamma/\gamma}$ corresponds to a massive tadpole, for the fourth type
we obtain a product of two massive tadpoles, while for the fifth type we
get a two-loop massive vacuum integral (with three internal lines).

The Taylor expansion operator $\cal{T}$ expands the denominators of
the integrand in the following way:
\be
\label{Tm}
{\cal{T}}_m \; \frac{1}{{[p^2 - m^2]}^{\nu}}
= \frac{1}{{[p^2]}^{\nu}} \; \sum_{j=0}^{\infty} \frac{(\nu)_j}{j!} \;
           \left( \frac{m^2}{p^2} \right)^j \; ,
\ee
\be
\label{Tmq}
{\cal{T}}_{m; q} \; \frac{1}{{[(k-q)^2 - m^2]}^{\nu}}
= \frac{1}{{[k^2]}^{\nu}} \; \sum_{j=0}^{\infty} \frac{(\nu)_j}{j!} \;
           \left( \frac{2 (kq) - q^2 + m^2}{k^2} \right)^j \; ,
\ee
where
\be
\label{poch}
(\nu)_j \equiv \frac{\Gamma(\nu+j)}{\Gamma(\nu)}
\ee
is the Pochhammer symbol . In fact, on the
r.h.s. of (\ref{Tmq}) it is understood that the real expansion goes
over the total power of $m$ and $q$ (so, for example, the terms $q^2,
\; m^2$ and $(kq)^2/k^2$ should be considered together). The reason is
that when we evaluate the massive vacuum integrals $J_{\Gamma/\gamma}$ with
these momenta $q$ in the numerator, the powers of
$q$ will be transformed into powers of the masses. If we have several
denominators to expand, then we should also collect all terms with the
same total power of masses and ``small'' momenta $q_i$.

Now we are able to consider what integrals correspond to different terms
of the asymptotic expansion (see Fig.2).

1. In this case, $\gamma=\Gamma$. All denominators of (\ref{defJ})
should be expanded in masses,
\be
\label{1term}
{\cal{T}}_{\{m_i\}} \; J(\{\nu_i\} ; \{m_i\} ; k)
= \sum_{j_1, \ldots , j_5 = 0}^{\infty}
\frac{(\nu_1)_{j_1} \ldots (\nu_5)_{j_5} }{j_1 ! \ldots j_5 !}
(m_1)^{j_1} \ldots (m_5)^{j_5} \; J(\{\nu_i + j_i\} ; \{0\} ; k) .
\ee
Note that if we consider the case $\nu_1= \ldots = \nu_5 =1$, the first
term of the expansion (\ref{1term}) (with $j_1 = \ldots = j_5 = 0$)
gives the well-known result:
$ - 6 \zeta(3) \pi^4/k^2 $. Two-loop massless integrals with higher integer
powers of denominators occurring on the r.h.s. of (\ref{1term}) can be
evaluated by use of the integration-by-parts technique
\cite{IBP} (see Appendix A).

2. Let us consider only the first contribution of the second type (see
Fig.2), when $\gamma$ is obtained from $\Gamma$ by removing line 1.
Then we get
\be
\label{2term}
\! \int \! \frac{\mbox{d}^n p}{{[p^2 \! - \! m_1^2]}^{\nu_1}} \;
{\cal{T}}_{m_2, \ldots , m_5; p}
\int \! \frac{\mbox{d}^n q}
   {{[q^2 \! - \! m_2^2]}^{\nu_2} {[(p\!-\!q)^2\! -\! m_3^2]}^{\nu_3}
     {[(k\!-\!p)^2 \!- \!m_4^2]}^{\nu_4}  {[(k\!-\!q)^2\! - \!m_5^2]}^{\nu_5} }
{}.
\ee
After expanding the integrand of the $q$-integral in masses and $p$, we
obtain products of massless one-loop integrals (see Appendix A) and
massive tadpoles with numerators that can be calculated by use
of eq.~(\ref{red1}) (see Appendix B). Other contributions
of the second type can be evaluated in the same way.

3. In the case when the central line is removed from $\gamma$, we obtain:
\bea
\label{3term}
\! \int \! \frac{\mbox{d}^n p}{{[p^2 \! - \! m_3^2]}^{\nu_3}} \;
\hspace{12cm} \nonumber \\ \times
{\cal{T}}_{m_1, m_2, m_4 , m_5; p}
\int \! \frac{\mbox{d}^n q}
   {{[(k\!+\! p \! - \!q)^2 \! - \! m_1^2]}^{\nu_1}
    {[(k\!-\!q)^2\! -\! m_2^2]}^{\nu_2}
     {[(p\!-\!q)^2 \!- \!m_4^2]}^{\nu_4}  {[q^2\! - \!m_5^2]}^{\nu_5} } .
\eea
After expansion, we obtain integrals of the same type as in the previous case.

4. There are no loop integrations in the subgraph $\gamma$, and we get
for the first contribution of the fourth type:
\bea
\label{4term}
\! \int \! \int \! \frac{\mbox{d}^n p}{{[p^2 \! - \! m_1^2]}^{\nu_1}}
                   \frac{\mbox{d}^n q}{{[q^2 \! - \! m_5^2]}^{\nu_5}} \;
\hspace{9cm} \nonumber \\ \times
{\cal{T}}_{m_2, m_3, m_4 ; p, q}
\left( \frac{1}{ {[(k\!-\!q)^2\! -\! m_2^2]}^{\nu_2}
 {[(k\! -\!p\!-\!q)^2\! -\! m_3^2]}^{\nu_3}
 {[(k\!-\!p)^2 \!- \!m_4^2]}^{\nu_4} } \right) .
\eea
As a result, we obtain products of two one-loop tadpoles with numerators
(also for the second contribution),
which can be evaluated by use of eq.~(\ref{red2}) (see Appendix B).

5. In this case we obtain a non-trivial two-loop vacuum
integral. For example, the first contribution of the fifth type
gives:
\bea
\label{5term}
 \int \! \int \! \frac{\mbox{d}^n p \; \mbox{d}^n q}
          {{[p^2 \! - \! m_1^2]}^{\nu_1} {[q^2 \! - \! m_2^2]}^{\nu_2}
               {[(p-q)^2 \! - \! m_3^2]}^{\nu_3}}
\hspace{5cm} \nonumber \\ \times
{\cal{T}}_{m_4, m_5; p, q}
\left( \frac{1} {{[(k\!-\!p)^2 \!- \!m_4^2]}^{\nu_4}
        {[(k\!-\!q)^2\! -\! m_5^2]}^{\nu_5}} \right) .
\eea
Expanding the denominators, we obtain two-loop vacuum integrals with
numerators, that can be evaluated by use of eq.~(\ref{red3})
presented in Appendix B
(the same for the second contribution of the fifth type). Note, that here
we obtain the same two-loop vacuum integrals as those involved in the
small-$k^2$ expansion (see \cite{DT}). In particular, for unit
powers of denominators the dependence on masses can be expressed
in terms of dilogarithms (see (\ref{I111})--(\ref{eq:Phi})).

So, the total asymptotic expansion is the sum of all terms presented, and we
see that all integrals corresponding to the coefficients of the expansion
can be evaluated.
\section{Analytical results}

In principle, eqs.~(\ref{1term})--(\ref{5term})
presented in the previous section (together with the formulae
of Appendices A and B) enable one to construct analytical
expressions for the coefficients of the large-$k^2$ expansion.
However, in the general case of unequal masses the higher-order
coefficients become rather cumbersome. To calculate these
coefficients, we used the {\sf REDUCE} system \cite{red}.
The algorithm constructed is applicable to integrals with
arbitrary values of masses, space-time dimension and (integer)
powers of denominators. If we are interested in the result
near $n=4$, we perform an expansion in $\ep = (4-n)/2$ to get
the divergent and finite parts of the coefficients.

One of the most important examples is the ``master''
two-loop diagram (presented in Fig.1a) in the case
$\nu_1= \ldots = \nu_5 = 1$. In this case, the result should
be finite as $n~\to~4$ (and it is a non-trivial check of the
algorithm that all the divergent contributions from separate
terms of (\ref{1term})--(\ref{5term}) cancel in this sum !).
A rigorous proof of the finiteness of the expansion was given
in \cite{MPI} (it was based on the so-called $R^*$-operation
\cite{R*}).

Let us define
\bea
\label{defMj}
\left. J(1, \ldots ,1; m_1, \ldots , m_5; k) \right|_{n=4}
& \equiv &  J( m_1, \ldots , m_5; k)
\hspace{3cm} \nonumber \\[2mm]
& \equiv & - \frac{\pi^4}{k^2}
{\cal M}(m_1, \ldots , m_5; k)
\equiv  - \frac{\pi^4}{k^2}
\sum_{j=0}^{\infty} \frac{{\cal M}_j}{(k^2)^j} , \hspace{1cm}
\eea
where the coefficient functions ${\cal M}_j$ include the powers
of masses and the logarithms of masses and momentum squared.
It is easy to see that the only integral contributing to
${\cal M}_0$ is $J^{(0)}(1,1,1,1,1)$  in
(\ref{1term}) (see eq.~(\ref{6zeta3}) in Appendix A).
So we get the obvious result that the expansion starts from
\be
\label{M0}
{\cal M}_0 = 6 \zeta(3) .
\ee

The ${\cal M}_1$ term already includes contributions of all
terms (\ref{1term})--(\ref{5term}) (with the exception of
(\ref{4term}) that begins to contribute starting from
${\cal M}_2$); this yields
\bea
\label{M1}
{\cal M}_1 &=& \frac{m_1^2}{2}
\left\{ \ln^2 \left( -\frac{k^2}{m_1^2} \right)
+ 4 \ln \left( -\frac{k^2}{m_1^2} \right)
- \ln \frac{m_2^2}{m_1^2} \; \ln \frac{m_3^2}{m_1^2} + 6
\right\}
\nonumber \\[2mm]
&&+ \left\{ \mbox{analogous terms with} \; m_2^2, \; m_4^2, \; m_5^2 \right\}
\nonumber \\[2mm]
&&+ \frac{m_3^2}{2}
\left\{ 2 \ln^2 \left( -\frac{k^2}{m_3^2} \right)
+ 4 \ln \left( -\frac{k^2}{m_3^2} \right)
- \ln \frac{m_1^2}{m_3^2} \; \ln \frac{m_2^2}{m_3^2}
- \ln \frac{m_4^2}{m_3^2} \; \ln \frac{m_5^2}{m_3^2}
\right\}
\nonumber \\[2mm]
&&+ \frac{1}{2} \left\{ F (m_1^2, m_2^2, m_3^2) + F (m_4^2, m_5^2, m_3^2)
\right\} \, ,
\eea
where the symmetric function $F$ is defined by (\ref{defF})--(\ref{eq:Phi})
(note that $F$ has the dimension of mass squared). We see that ${\cal M}_1$
contains also $\ln(-k^2)$ and $\ln^2 (-k^2)$ terms, which
means that our expansion (at $n=4$) is not a usual Taylor expansion.
In fact, the highest power of $\ln(-k^2)$ is connected with the
highest order of pole in $\ep$ (double pole) that can occur in the two-loop
integrals involved. Note that for positive $k^2$ the sign of the imaginary part
produced by these logarithms is defined by the ``causal'' $i0$-prescription,
\be
\ln(-k^2 - i0) = \ln(k^2) - i \pi \hspace{2cm} ( k^2 > 0 ).
\ee

Let us also present the result for the next term of the expansion,
\bea
\label{M2}
{\cal M}_2 &=& \frac{m_1^4}{8}
\left\{ 2 \ln^2 \left( -\frac{k^2}{m_1^2} \right)
       +4 \ln \left( -\frac{k^2}{m_1^2} \right)
       -2 \ln{\frac{m_2^2}{m_1^2}} \ln{\frac{m_3^2}{m_1^2}} +7 \right\}
 \nonumber \\[2mm]
&& \hspace{1cm}
+ \left\{ \mbox{analogous terms with} \; m_2^4, \; m_4^4, \; m_5^4 \right\}
 \nonumber \\[2mm]
&+& \frac{m_3^4}{4}
\left\{ -2 \ln^2 \left( -\frac{k^2}{m_3^2} \right)
       -2 \ln \left( -\frac{k^2}{m_3^2} \right)
       + \ln{\frac{m_1^2}{m_3^2}} \ln{\frac{m_2^2}{m_3^2}}
       + \ln{\frac{m_4^2}{m_3^2}} \ln{\frac{m_5^2}{m_3^2}}
       +6 \right\}
  \nonumber \\[2mm]
&-& \frac{1}{2} (m_1^2 m_2^2  + m_4^2 m_5^2 )
  \nonumber \\[2mm]
&+& \frac{m_1^2 m_4^2}{2}
\left\{  \ln^2 \left( -\frac{k^2}{m_1^2} \right)
        + \ln^2 \left( -\frac{k^2}{m_4^2} \right)
       +4 \ln \left( -\frac{k^2}{m_1^2} \right)
       +4 \ln \left( -\frac{k^2}{m_4^2} \right)
\right. \nonumber \\ && \left.
 \hspace{1cm} - \ln{\frac{m_2^2}{m_1^2}} \ln{\frac{m_3^2}{m_1^2}}
       - \ln{\frac{m_3^2}{m_4^2}} \ln{\frac{m_5^2}{m_4^2}}
       +8 \right\}
 \nonumber \\[2mm]
&&\hspace{1cm}
+ \left\{ \mbox{analogous term with} \; m_2^2 m_5^2 \right\}
 \nonumber \\[2mm]
&+& \frac{m_1^2 m_5^2}{2}
  \left\{  2 \ln^2 \left( -\frac{k^2}{m_1^2} \right)
        + 2 \ln^2 \left( -\frac{k^2}{m_5^2} \right)
       +2 \ln \left( -\frac{k^2}{m_1^2} \right)
       +2 \ln \left( -\frac{k^2}{m_5^2} \right)
\right. \nonumber \\ && \left. \hspace{1cm}
         - \ln{\frac{m_2^2}{m_1^2}} \ln{\frac{m_3^2}{m_1^2}}
       - \ln{\frac{m_3^2}{m_5^2}} \ln{\frac{m_4^2}{m_5^2}}
       - \ln^2 \frac{m_1^2}{m_5^2}  +2 \right\}
 \nonumber \\[2mm]
&&\hspace{1cm}
+ \left\{ \mbox{analogous term with} \; m_2^2 m_4^2 \right\}
 \nonumber \\[2mm]
&+& \frac{m_1^2 m_3^2}{2}
\left\{  2 \ln^2 \left( -\frac{k^2}{m_3^2} \right)
        - 2 \ln \left( -\frac{k^2}{m_1^2} \right)
       - \ln{\frac{m_1^2}{m_3^2}} \ln{\frac{m_2^2}{m_3^2}}
       - \ln{\frac{m_4^2}{m_3^2}} \ln{\frac{m_5^2}{m_3^2}}
       -8 \right\}
 \nonumber \\[2mm]
&&\hspace{1cm}
+ \left\{ \mbox{analogous terms with} \; m_2^2 m_3^2, m_4^2 m_3^2,
    m_5^2 m_3^2 \right\}
 \nonumber \\[2mm]
&+& \frac{1}{4}
\left\{ \left( m_1^2 + m_2^2 -m_3^2 + 2 m_4^2 + 2 m_5^2 \right) \;
                       F(m_1^2, m_2^2, m_3^2)
\right. \nonumber \\ && \left. \hspace{2cm}
+\left( 2 m_1^2 + 2 m_2^2 -m_3^2 +  m_4^2 +  m_5^2 \right) \;
                       F(m_4^2, m_5^2, m_3^2) \right\} .
\eea
Higher contributions are more cumbersome, and we do not present them here
(but we are going to use them below, when comparing our expansion
with the results of numerical integration). By use of the
{\sf REDUCE} system \cite{red}, for the general massive case of the
integral (\ref{defMj}) (when all five masses are arbitrary)
we obtained analytical results
for the coefficient functions up to ${\cal M}_6$.

There are also some other possibilities to check the correctness
of our results (in addition to cancellation of $1/\ep$ poles). For example,
in ref.~\cite{Bro} analogous results were presented
for some special cases when some of the masses are zero while
others are equal (see also ref.~\cite{BFT} where these results
were generalised to the case of arbitrary space-time dimension $n$).
In these cases, we compared our results for the
coefficients (at $n=4$) and found complete agreement.
\section{Numerical results}
In this section, we will continue to focus on the ``master'' two-loop
integral corresponding to Fig.~1a. In general, it has two two-particle
thresholds, at $k^2=(m_1+m_4)^2$ and $k^2=(m_2+m_5)^2$, and two
three-particle thresholds at $k^2=(m_1+m_3+m_5)^2$ and
$k^2=(m_2+m_3+m_4)^2$. The asymptotic expansion theorem quoted
in (\ref{theorem}) provides a series of approximations to this integral
of the form (see eq.~(\ref{defMj})):
\be
\label{eq:defMN}
{\cal M}^{(N)} \equiv \sum_{j=0}^{N} \frac{{\cal M}_j}{{(k^2)}^j} \, ,
\ee
such that the remainder behaves like
\be
{\cal M} - {\cal M}^{(N)} = {\cal O} \left(
          {(k^2)^{-N-1} \ln^2 (-k^2)} \right)
\ee
as $ k^2 \to \infty$.

Strictly speaking, this does not imply convergence of the series
for any fixed value of $k^2$, but from experience with special
cases where exact analytical results are known, one would expect
the series to converge when $|k^2|$ is larger than the highest
threshold. In order to see whether this is true in the general
case, and whether the asymptotic expansion can be used as a
practical means of calculating these integrals, we made some
numerical comparisons for two physical examples.

The first example is:
\be
\label{eq:example1}
J(M_W,M_W,M_Z,m_b,m_b;k),
\ee
where the $M_W$, $M_Z$ and $m_b$ denote the masses of the
$W$-boson, the $Z$-boson and the bottom quark. This integral is a
contribution to the top quark self-energy.  In this case,
both the two-particle thresholds coincide. The three-particle
thresholds also coincide with each other.

The other example occurs in the photon, the Higgs and the Z-boson
self-energies:
\be
\label{eq:example2}
J(m_t,m_t,M_Z,m_t,m_t;k).
\ee
Here $m_t$ is the top quark mass. As in the first example, there are
only two distinct thresholds.  We calculated (\ref{eq:example1}) and
(\ref{eq:example2}) numerically using the method of ref.~\cite{Kre}.
The values we took for the masses were:
\be
\begin{array}{cccc}
M_Z=91 \, \mbox{GeV, }&
M_W=80 \, \mbox{GeV, }&
m_t=140\, \mbox{GeV, }&
m_b=5  \, \mbox{GeV. }
\end{array}
\ee
The results are displayed in Figs.~\ref{fig:WWZbb} and
\ref{fig:ttZtt}.  In each plot, the first threshold is exactly on the
left edge and the position of the highest threshold is indicated by a
dashed vertical line. The dotted horizontal line shows the lowest order
asymptotic approximation \mbox{${\cal M}^{(0)}\equiv{\cal M}_0=6
\zeta(3)=7.21234\dots\,$} (the imaginary part is zero to this order).
The curves
labeled $1,2,\ldots$ show the approximations ${\cal M}^{(1)}, {\cal
M}^{(2)},\ldots$ defined by eq.~(\ref{eq:defMN}). The results of the
numerical integration program are shown as crosses.

At large values of $k^2$, say ten times the highest threshold or
higher, ${\cal M}^{(3)}$ already agrees with the numerical results
to within 0.01\%, which is the order of magnitude of the error
in the numerical results.

The most interesting region is immediately above the highest threshold,
where we would still expect the expansion to converge, but more slowly
than at large $k^2$. In fact, in our first example (\ref{eq:example1}),
${\cal M}^{(4)}$ is still accurate to within 1\% on the threshold itself.
In the second example (\ref{eq:example2}), the convergence near the
threshold is less rapid. This region is shown in more detail
in Fig.~\ref{fig:closeup}, where the left edge of the plots corresponds
to the highest threshold. On the threshold itself, the error
in ${\cal M}^{(6)}$ is about 6\% (in the real part), but it drops
down to less than 1\% by the time $k^2$ is a factor of 1.2 above
the threshold.

Note that in \cite{KK} radiative corrections to the on-shell
$H~\to~b\bar{b}$ squared amplitude (represented by a diagram
like Fig.1a) were examined, and it was pointed out
that the mass correction ${\cal M}_1$ essentially improved
the description above the last threshold.

Finally, when we go below the highest threshold, we see large deviations
between the asymptotic approximations and the numerical values,
and the series ceases to converge. From this we see that
the asymptotic expansion can only be applied to the region above
the highest threshold, as expected.

\section{Conclusions}

Thus, in the present paper we  considered an algorithm
to construct an asymptotic expansion of two-loop self-energy diagrams
(see Fig.1) for large values of $k^2$. This algorithm can be applied to
the general case of different masses, and integer powers of
denominators. We used the asymptotic expansion theorem (\ref{theorem})
to derive the different contributions to the expansion
(see (\ref{1term})--(\ref{5term})). In any order, the coefficients
of the expansion can be expressed in terms of known one-
and two-loop propagator-type massless integrals, and one- and
two-loop massive vacuum integrals.

By use of the {\sf REDUCE} system, we wrote a program that automatically
generates analytical expressions for the coefficients of the asymptotic
expansion.
Then we considered the ``master'' two-loop diagram with different masses
(Fig.1a), and we obtained expressions for the terms of the asymptotic expansion
of the integral (\ref{defMj}) up to $1/(k^2)^7$ terms (with logarithms).
The first three coefficient functions ${\cal M}_{0,1,2}$ are given by equations
(\ref{M0}), (\ref{M1}), (\ref{M2}). For some concrete diagrams
occurring in the Standard Model, we made a numerical comparison
with the results of a two-fold numerical integration based on
the algorithm of \cite{Kre}. We found that in the region above the highest
physical threshold of the diagram, our expansion converges well,
and in the region not very close to the threshold it is sufficient to
take only a few terms of the expansion to provide a good accuracy.

Thus, the present paper (together with \cite{DT}) solves the
problem of constructing asymptotic expansions of massive two-loop self-energy
diagrams in the limits of small and large $k^2$ values (when
we are either below the lowest physical threshold or above
the highest one). The analytical description of the behaviour
between the thresholds still remains a problem.
Note, however, that the general asymptotic expansion
theorem can also be used for other limits in order to
calculate the considered diagram in situations when some of the masses
are large and the others are small.

A procedure analogous to the one considered here can be also applied to
three-point two-loop vertex diagrams, when all external lines
are above the corresponding thresholds (and also in some other cases).
Note that some two-loop massless integrals with off-shell
external momenta (needed for such an expansion) were calculated
in ref.~\cite{UD}. For numerical comparison, a
parametric integral representation obtained in ref.~\cite{Kre2}
can be used.

\section*{Acknowledgements}
We are grateful to F.A.~Berends and W.L.~van~Neerven
for their assistance and helpful discussions.
We are also grateful to D.J.~Broadhurst, R.~Buchert, A.L.~Kataev
and D.~Kreimer for useful discussions, and to H.~Kuijf and
S.C.~van~der~Marck for writing the program that calculates the integral
(\ref{defMj}) numerically.
One of the authors (A.D.) is grateful to Instituut--Lorentz for
their hospitality and to the Nederlandse organisatie voor
Wetenschappelijk Onderzoek (NWO) for financial support.
Another author (V.S.) is
thankful to the Institute for Theoretical
Physics of G\"ottingen University for kind hospitality and to
the Heraeus Foundation for support.

\newpage

{\large\bf Appendix A: Massless integrals } \\[1ex]
Here we present some formulae needed for evaluation of
one- and two-loop massless integrals occurring in the
paper. These results are well known, and we write them
only for completeness.

The massless one-loop integral is
\bea
\label{defI0}
I^{(0)} (\nu_1, \nu_2) \equiv
\int \frac {\mbox{d}^n p}{[p^2]^{\nu_1} [(k-p)^2]^{\nu_2}}
\hspace{8cm}
\nonumber \\
= i^{1-n} \pi^{n/2}  (k^2)^{n/2-\nu_1-\nu_2}
\frac{\Gamma(n/2-\nu_1) \Gamma(n/2-\nu_2) \Gamma(\nu_1+\nu_2-n/2)}
     {\Gamma(\nu_1) \Gamma(\nu_2) \Gamma(n-\nu_1-\nu_2)} ,
\eea
where $n=4-2\ep$ is the space-time dimension.

Massless two-loop integrals are defined by (see (\ref{defJ}))
\be
\label{defJ0}
J^{(0)}(\nu_1,\nu_2,\nu_3,\nu_4,\nu_5)
\equiv J(\{\nu_i\}; \{0\}; k)
\ee
By use of symmetry properties of the diagram in Fig.1a, and the
integration by parts formula \cite{IBP},
\bea
\label{IBP}
J^{(0)} (\nu_1, \nu_2, \nu_3, \nu_4, \nu_5)
= (\nu_1 + 2 \nu_3 + \nu_4 -n)^{-1} \hspace{5cm} \nonumber \\
\times \left\{ \nu_1 \left[
J^{(0)} (\nu_1 +1, \nu_2 -1, \nu_3, \nu_4, \nu_5)
- J^{(0)} (\nu_1 +1, \nu_2, \nu_3 -1, \nu_4, \nu_5)
\right] \right. \nonumber \\
\hspace*{4mm} \left. + \nu_4 \left[
J^{(0)} (\nu_1, \nu_2, \nu_3, \nu_4 +1, \nu_5 -1)
- J^{(0)} (\nu_1, \nu_2, \nu_3 -1, \nu_4 +1, \nu_5)
\right] \right\} ,
\eea
these integrals (with positive integer $\nu$'s) can be reduced to
the following ``boundary'' integrals:
\be
\label{bound1}
J^{(0)} (\nu_1, \nu_2, 0, \nu_4, \nu_5)
= I^{(0)} (\nu_1, \nu_4) \; I^{(0)} (\nu_2, \nu_5) ;
\ee
\be
\label{bound2}
J^{(0)} (\nu_1, \nu_2, \nu_3, \nu_4, 0)= (k^2)^{\nu_2+\nu_3-n/2} \;
 I^{(0)} (\nu_2, \nu_3) \; I^{(0)} (\nu_1+\nu_2+\nu_3-n/2, \nu_4) .
\ee
For example,
\be
\label{6zeta3}
J^{(0)} (1,1,1,1,1) = - \frac{\pi^4}{k^2} \; 6 \zeta(3) + {\cal O}(\ep)
\ee
(note that $\zeta(3)$ does not occur in the divergent and finite parts of
any other integrals $J^{(0)}$ with positive $\nu$'s).

\vspace{8mm}

{\large\bf Appendix B: Massive vacuum integrals } \\[1ex]
The result for one-loop massive tadpole integral is well known
\cite{tHV-72} :
\be
\label{defK}
K(\nu,m) \equiv \int \frac{\mbox{d}^n p}{[p^2-m^2]^{\nu}}
= i^{1-n} \pi^{n/2} (-m^2)^{n/2-\nu} \frac{\Gamma(\nu-n/2)}{\Gamma(\nu)} .
\ee

Two-loop vacuum massive integrals
\be
\label{defI}
I(\nu_1,\nu_2,\nu_3; m_1, m_2, m_3) \equiv
\int \int \frac{\mbox{d}^n p \; \mbox{d}^n q}
   {[p^2-m_1^2]^{\nu_1} [q^2-m_2^2]^{\nu_2}
    [(p-q)^2-m_3^2]^{\nu_3}}
\ee
have been studied e.g. in ref.~\cite{DT} (some special cases were
also considered in ref.~\cite{walnut}). For example,
for unit powers of denominators we get
\bea
\label{I111}
I (1,1,1; m_1, m_2, m_3) = \pi^{4 -2\ep} \Gamma^2 (1+\ep) \;
   (1 + 3\ep + 7\ep^2)
\hspace{3.7cm} \nonumber \\
\times \left\{- \frac{1}{2 \ep^2} (m_1^2 + m_2^2 +m_3^2)
+ \frac{1}{\ep} (m_1^2 \ln{m_1^2} +m_2^2 \ln{m_2^2} +m_3^2
\ln{m_3^2}) \right. \nonumber \\
-\frac{1}{2} \left[ m_1^2 \ln^2 m_1^2 + m_2^2 \ln^2 m_2^2 + m_3^2 \ln^2 m_3^2
\right. \hspace{4cm} \nonumber \\
+ (m_1^2 +m_2^2 -m_3^2) \ln{m_1^2} \ln{m_2^2}
+ (m_1^2 -m_2^2 +m_3^2) \ln{m_1^2} \ln{m_3^2}
 \nonumber \\ \left. \left.
+ ( -m_1^2 +m_2^2 +m_3^2) \ln{m_2^2} \ln{m_3^2}
+ F (m_1^2, m_2^2, m_3^2) \right] \right\}  + {\cal{O}}(\ep)
\eea
where the function $F$ is symmetric with respect to $m_1^2, m_2^2, m_3^2$,
and is defined by (see \cite{DT})
\be
\label{defF}
F (m_1^2, m_2^2, m_3^2) \equiv m_3^2
\lambda^2  \left( x, y \right)
\Phi \left( x, y \right)
\ee
with
\be
\label{xy}
x \equiv \frac{m_1^2}{m_3^2} \hspace{1cm} , \hspace{1cm}
y \equiv \frac{m_2^2}{m_3^2} ,
\ee
\be
\label{lambda}
\lambda^2 (x,y) = (1-x-y)^2 - 4 x y ,
\ee
\begin{eqnarray}
\label{eq:Phi}
\Phi(x,y)&=& \frac{1}{\lambda} \left\{
          2 \ln \left( \frac {1+x-y-\lambda}{2} \right)
                    \ln \left( \frac {1-x+y-\lambda}{2} \right)
           - \ln x \ln y    \right.
\nonumber \\   [0.3 cm] &&  \left.
 -2 \, \Li{2}{ \frac {1+x-y-\lambda}{2} }
         -2 \, \Li{2}{ \frac {1-x+y-\lambda}{2} }
  + \frac {\pi^2}{3} \right\} \, .
\end{eqnarray}
If the largest mass is $m_1$ or $m_2$, we should choose this mass
as the dimensionless-making parameter in (\ref{defF}) and (\ref{xy})
(instead of $m_3$).
In the region where $\lambda^2 < 0$, the function (\ref{defF})
can be represented in terms of Clausen's functions (see \cite{DT}).
To obtain results for integrals (\ref{defI}) with higher integer
powers of denominators, a recursive procedure based on integration
by parts \cite{IBP} can be applied (see in \cite{DT}).
It is interesting that the same function (\ref{eq:Phi}) also occurs
when we evaluate massless triangle diagrams (see e.g. in \cite{JPA};
in this case $x$ and $y$ are dimensionless ratios of external
momenta squared).

We also need to evaluate massive integrals with numerators. To do
this, in the one-loop case we used the following explicit formula:
\bea
\label{red1}
\left. \int \frac{\mbox{d}^n p}{[p^2-m^2]^{\nu}}
\; [2(k_1 p)]^{N_1} \; [2(k_2 p)]^{N_2}
\right|_{N_1+N_2  \mbox{-- \small even}}
\hspace{5cm} \nonumber \\[3mm]
= \frac{N_1 ! \; N_2 !}{{(n/2)}_{(N_1+N_2)/2}}
\left\{
\begin{array}{c} {} \\ {} \\ \sum \\ {}_{2j_1+j_3=N_1} \\
                {}_{2j_2+j_3=N_2}  \end{array}
\frac{{(k_1^2)}^{j_1} {(k_2^2)}^{j_2} {[2(k_1 k_2)]}^{j_3}}
     {j_1 ! \; j_2 ! \; j_3 !} \right\}
 \int \frac{\mbox{d}^n p}{[p^2-m^2]^{\nu}} \;
 (p^2)^{(N_1 +N_2)/2},
\eea
and the integral on the l.h.s. vanishes if $(N_1+N_2)$ is odd.
The sum in braces goes
over all non-negative integers $j_1, j_2, j_3$ obeying two conditions:
$2j_1+j_3=N_1 \;$ and $ \;2j_2+j_3=N_2$ (so, in fact it
is a one-fold finite sum). We also use a standard notation for the
Pochhammer symbol (\ref{poch}).

When we have a two-loop integral with a numerator and two denominators
(corresponding to a product of two one-loop tadpoles), the
following formula can be derived:
\bea
\label{red2}
\left. \int \int \frac{\mbox{d}^n p \; \mbox{d}^n q}
                 {[p^2-m_1^2]^{\nu_1} [q^2-m_2^2]^{\nu_2}}
\; [2(k p)]^{N_1} \; [2(k q)]^{N_2} \; [2(p q)]^{N_3}
\right|_{ \begin{array}{c}
    {}_{N_1+N_3  \mbox{-- \small even}} \\[-1mm]
    {}_{N_2+N_3  \mbox{-- \small even}}
         \end{array} }
\hspace{1cm} \nonumber \\[3mm]
= \frac{N_1 !\; N_2 !\; {(k^2)}^{(N_1 +N_2)/2}}{{(n/2)}_{(N_1+N_2)/2}}
\left\{
\begin{array}{c} {} \\ {} \\ \sum \\ {}_{2j_1+j_3=N_1} \\
                {}_{2j_2+j_3=N_2}  \end{array}
\frac{(N_3+ j_3)!}
     {j_1 !\; j_2 !\; j_3 !\; ((N_3+j_3)/2)!\; {(n/2)}_{(N_3+j_3)/2}}
\right\}
\nonumber \\[3mm]
 \times \int \int \frac{\mbox{d}^n p \; \mbox{d}^n q}
                 {[p^2-m_1^2]^{\nu_1} [q^2-m_2^2]^{\nu_2}}
\; {(p^2)}^{(N_1 + N_3)/2} \; {(q^2)}^{(N_2 + N_3)/2} ,
\eea
and the integral on the l.h.s. of (\ref{red2}) is equal to zero
if $(N_1+N_3)$ or $(N_2+N_3)$ is odd. The sum in braces is of the same
structure as in (\ref{red1}).

We also need an analogous formula for the two-loop integral with three
denominators:
\bea
\label{red3}
\left.
\int \int \frac{\mbox{d}^n p \; \mbox{d}^n q}
   {[p^2-m_1^2]^{\nu_1} [q^2-m_2^2]^{\nu_2}
    [(p-q)^2-m_3^2]^{\nu_3}}
\; [2(k p)]^{N_1} \; [2(k q)]^{N_2}
\right|_{N_1+N_2  \mbox{-- \small even}}
\hspace{0.5cm} \nonumber \\[3mm]
= \frac{N_1 !\; N_2 !\; {(k^2)}^{(N_1 +N_2)/2}}{{(n/2)}_{(N_1+N_2)/2}}
\begin{array}{c} {} \\ {} \\ \sum \\ {}_{2j_1+j_3=N_1} \\
                {}_{2j_2+j_3=N_2}  \end{array}
\frac{1}{j_1 !\; j_2 !\; j_3 !}
\hspace{5cm} \nonumber \\[3mm]
\! \! \! \! \! \! \! \! \! \! \! \! \hspace*{-2cm}
\times \int \int \frac{\mbox{d}^n p \; \mbox{d}^n q}
   {[p^2\!-\!m_1^2]^{\nu_1} [q^2\!-\!m_2^2]^{\nu_2}
    [(p-q)^2\!-\!m_3^2]^{\nu_3}} \;
{(p^2)}^{j_1} {(q^2)}^{j_2} {[2(p q)]}^{j_3} ,\hspace{3mm}
\eea
and this integral is also equal to zero if $(N_1 + N_2)$ is odd.

In all formulae (\ref{red1}), (\ref{red2}) and (\ref{red3}),
the remaining momenta in the numerators on the r.h.s. can be
expressed in terms of denominators, and we arrive at the
result expressed in terms of integrals (\ref{defK}) and
(\ref{defI}) without numerators.

\newpage


\newpage

\section*{Figure captions}
\begin{figure}[htb]
\caption[]{The real and imaginary parts of
           $J(M_W,M_W,M_Z,m_b,m_b;k)$ for \mbox{$k^2 \geq (M_W+m_b)^2$}.
            The second threshold is at \mbox{$(M_W+M_Z+m_b)^2$}.}
\label{fig:WWZbb}
\end{figure}

\begin{figure}[htb]
\caption[]{The real and imaginary parts of
           $J(m_t,m_t,M_Z,m_t,m_t;k)$ for \mbox{$k^2 \geq 4 m_t^2$}.
           The second threshold is at \mbox{$(2m_t+M_Z)^2$}.}
\label{fig:ttZtt}
\end{figure}

\begin{figure}[htb]
\caption[]{The real and imaginary parts of
           $J(m_t,m_t,M_Z,m_t,m_t;k)$ for \mbox{$k^2 \geq (2 m_t + M_Z)^2$}.}
\label{fig:closeup}
\end{figure}

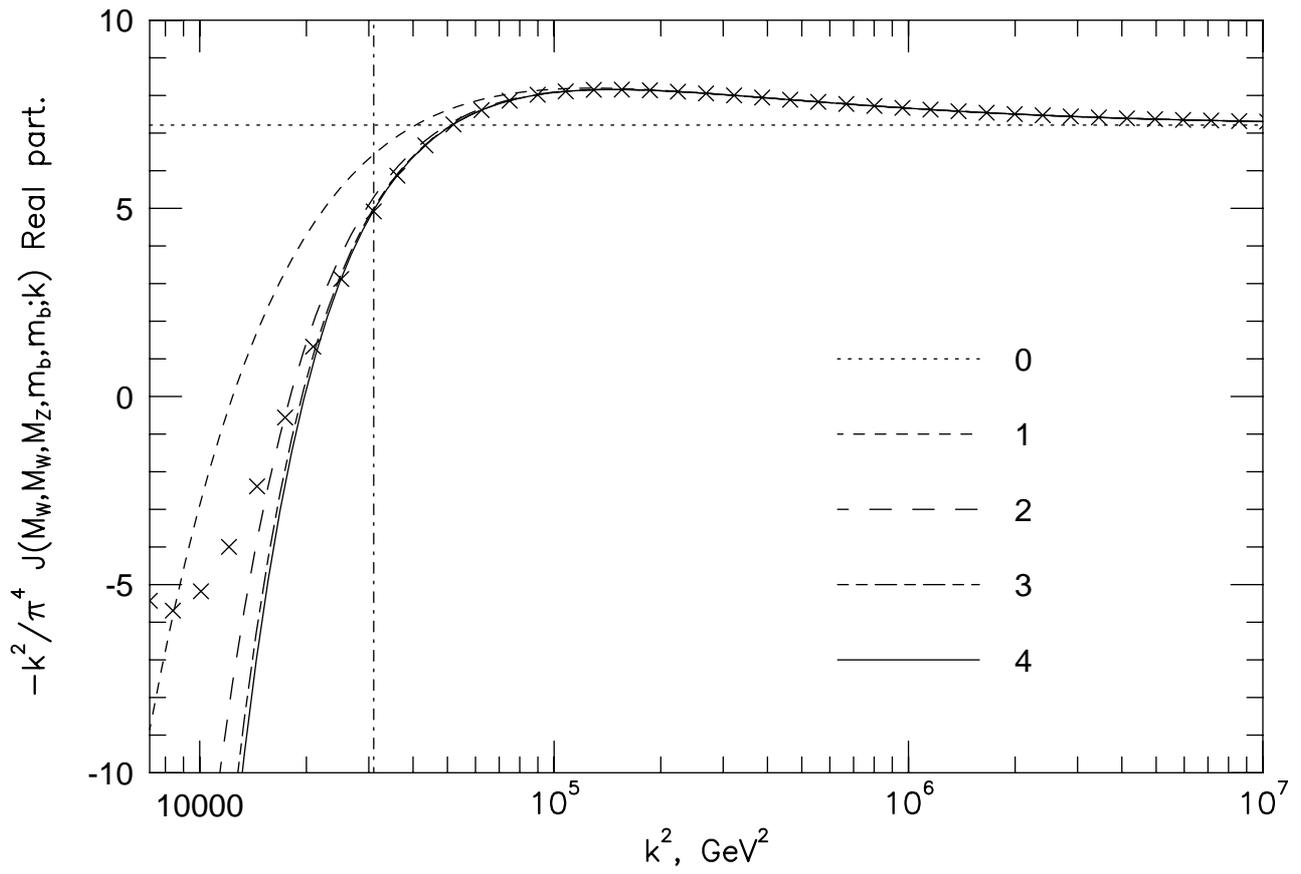
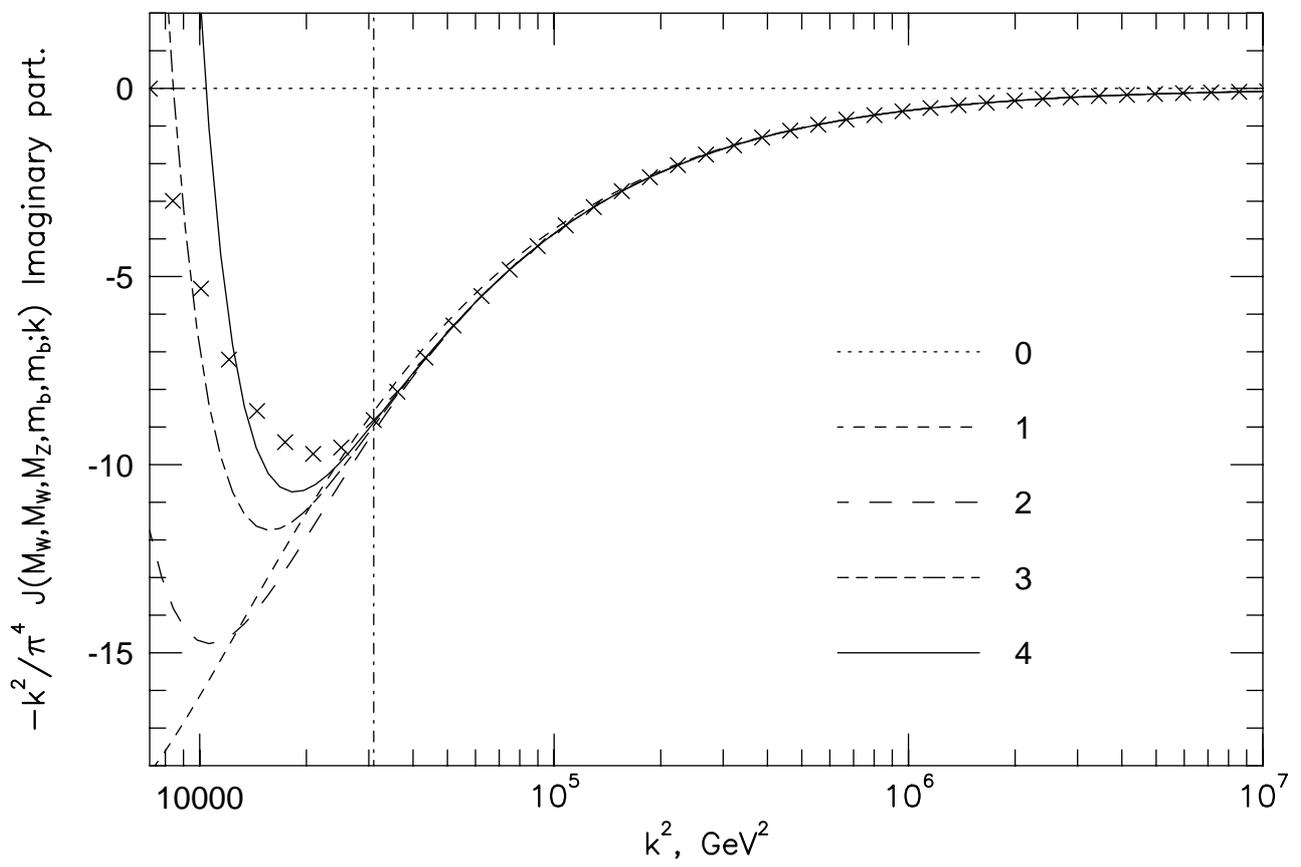
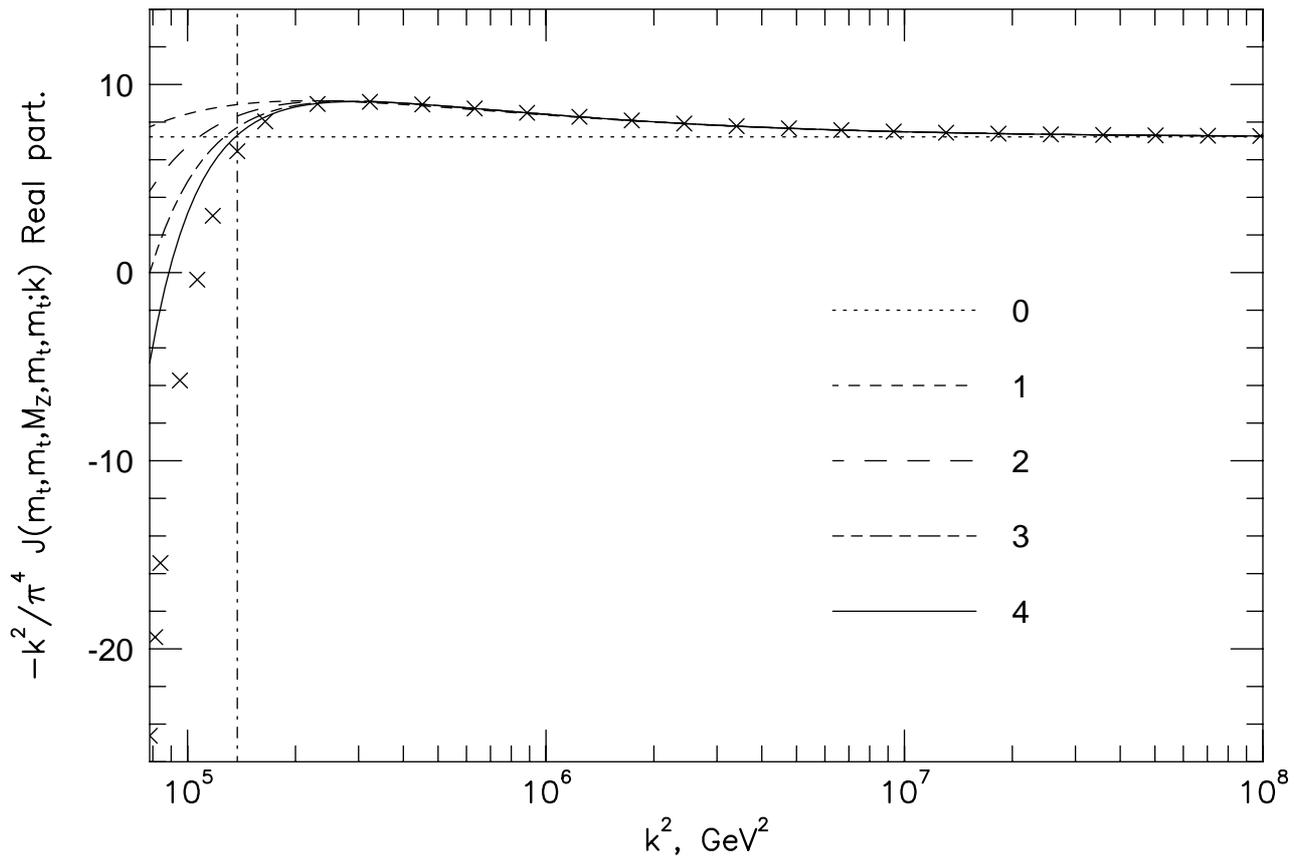
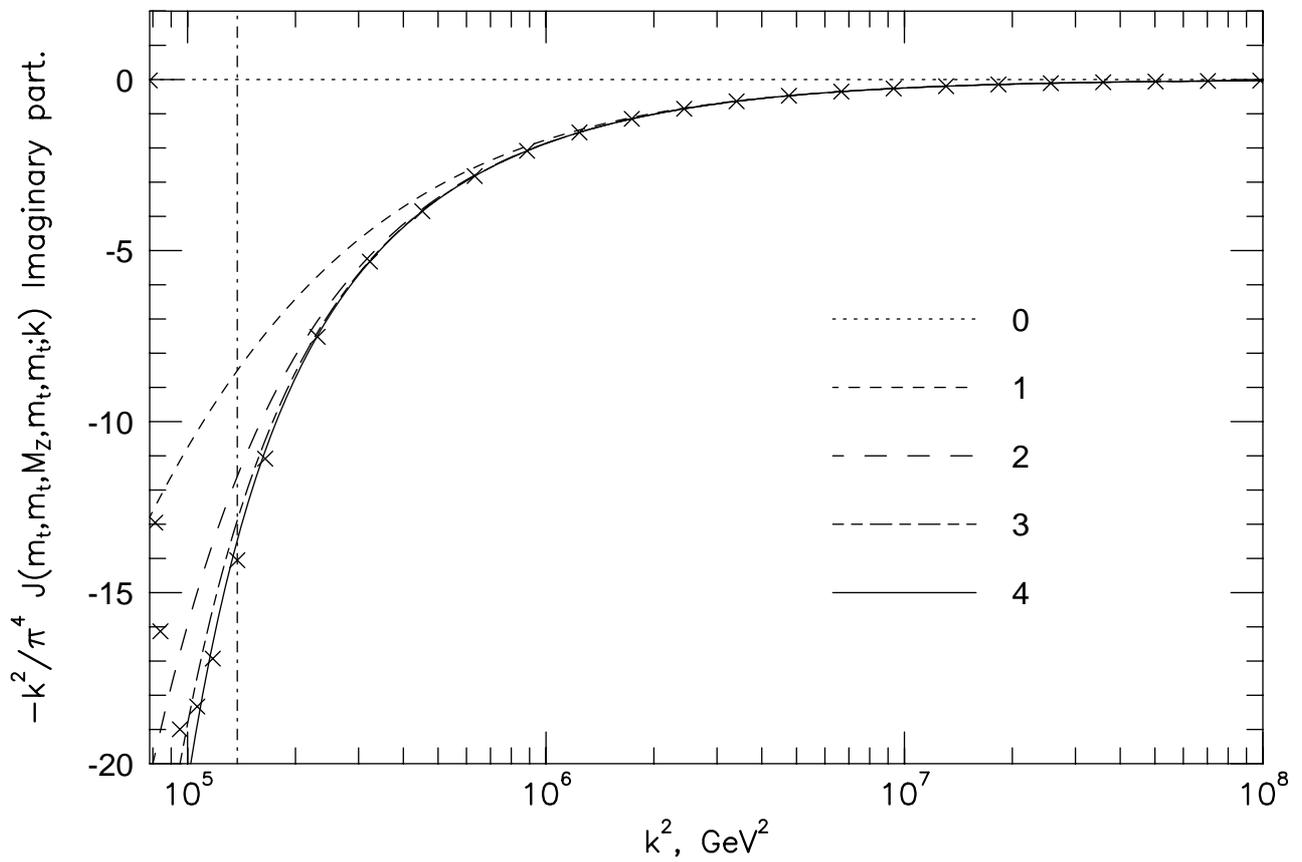
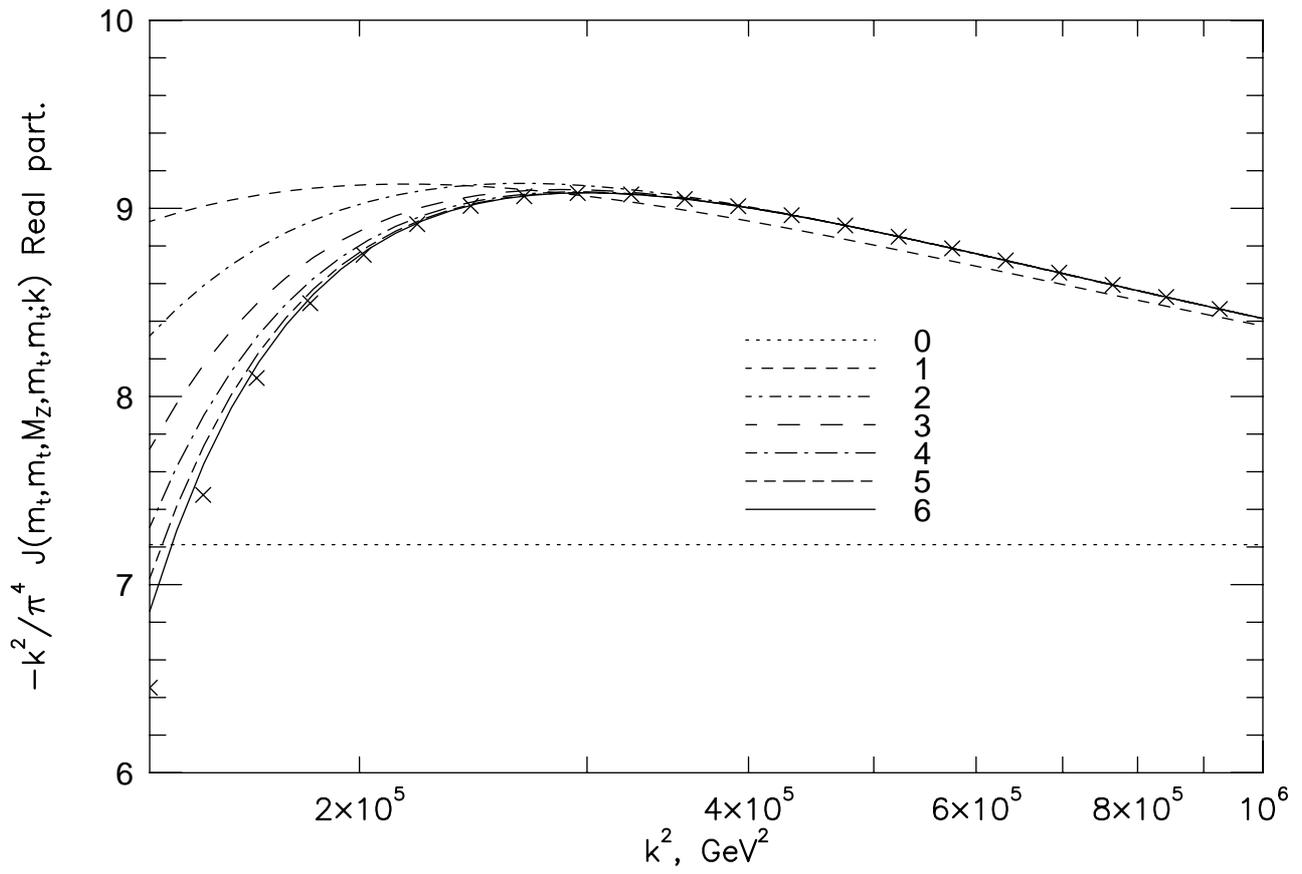
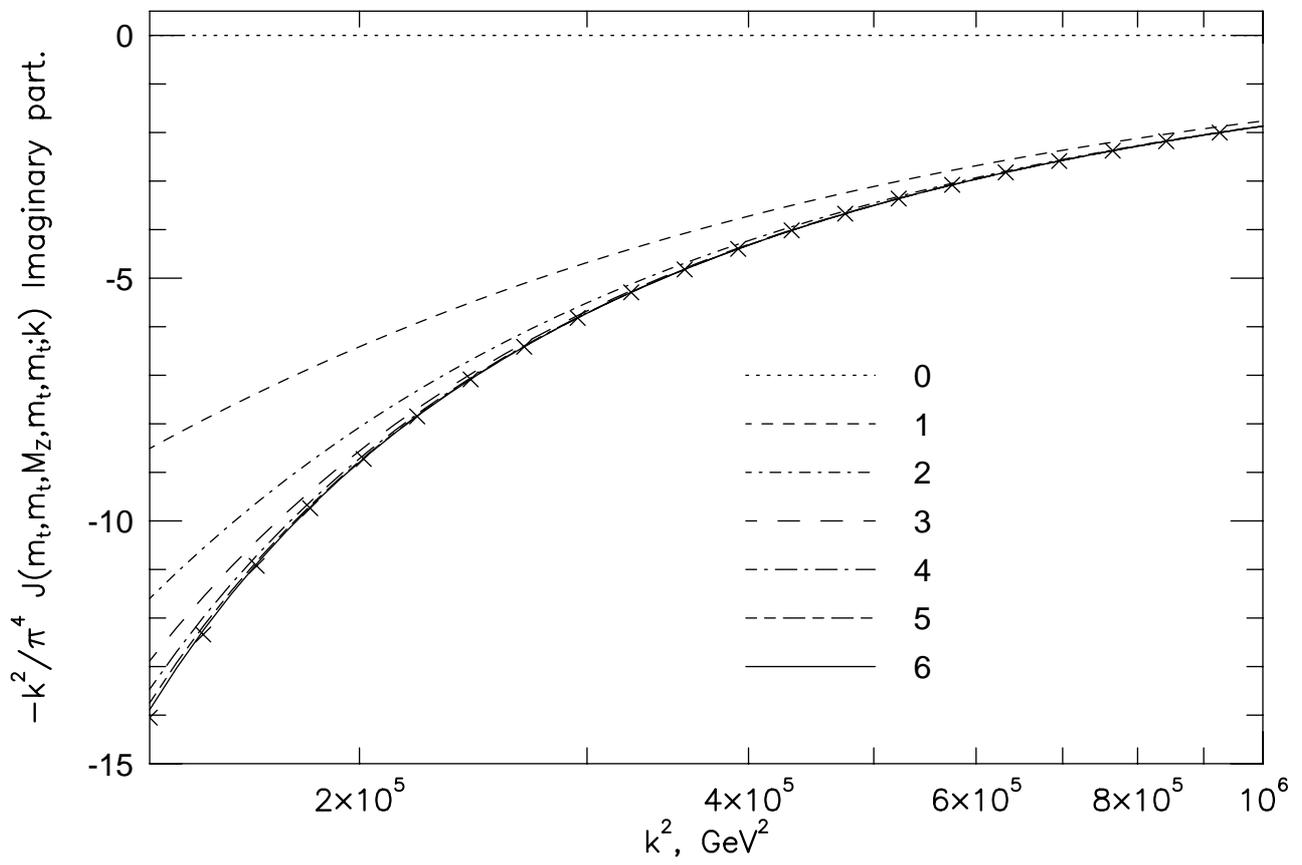
\end{document}